\documentclass[10pt,journal]{IEEEtran}
\usepackage{color}
\pdfoutput=1
\usepackage{cite,amsmath}
\usepackage[numbers,sort&compress]{natbib}
\usepackage{graphicx}
\usepackage{epstopdf}
\usepackage{subfigure}
\usepackage{stfloats}
\usepackage[justification=centering]{caption}
\usepackage{amsfonts,amssymb}
\usepackage{multirow}
\usepackage{algorithm}
\usepackage{algorithmicx}
\usepackage{algpseudocode}
\usepackage{array}
\usepackage[caption=false,font=normalsize,labelfont=sf,textfont=sf]{subfig}
\usepackage{textcomp}
\usepackage{verbatim}
\usepackage{graphicx}
\usepackage{cite}
\ifCLASSINFOpdf
\else

\fi

\begin{document}

\title{Variational Bayesian Learning Based Localization and Channel Reconstruction in RIS-aided Systems}

\author{Yunfei Li, Yiting Luo, Xianda Wu, Zheng Shi,  Shaodan Ma and Guanghua Yang
\thanks{Y. Li and Y. Luo are with the Department of Electrical Engineering, Anhui Polytechnic University, Wuhu City, China (email: lyf@mail.ahpu.edu.cn; lyt1222@ahpu.edu.cn).}
\thanks{X. Wu is with the School of Electronics and Information Engineering,
South China Normal University, Foshan 528000, China  (email: xiandawu@m.scnu.edu.cn).}
\thanks{G. Yang and Z. Shi are with the School of Intelligent Systems Science and Engineering, Jinan University, Zhuhai 519070, China (e-mails: zhengshi@jnu.edu.cn; ghyang@jnu.edu.cn).}
\thanks{S. Ma is with the State Key Laboratory of Internet of Things for Smart City and the Department
of Electrical and Computer Engineering, University of Macau, Taipa, Macao, China (e-mail: shaodanma@um.edu.mo).}
\vspace{-9mm}
}
\maketitle

\begin{abstract}
The emerging immersive and autonomous services have posed stringent requirements on both communications and localization. By considering the great potential of reconfigurable intelligent surface (RIS), this paper focuses on the joint channel estimation and localization for RIS-aided wireless systems. As opposed to existing works that treat channel estimation and localization independently, this paper exploits the intrinsic coupling and nonlinear relationships between the channel parameters and user location for enhancement of both localization and channel reconstruction. By noticing the non-convex, nonlinear objective function and the sparser angle pattern, a variational Bayesian learning-based framework is developed to jointly estimate the channel parameters and user location through leveraging an effective approximation of the posterior distribution. The proposed framework is capable of unifying near-field and far-field scenarios owing to exploitation of sparsity of the angular domain. Since the joint channel and location estimation problem has a closed-form solution in each iteration, our proposed iterative algorithm performs better than the conventional particle swarm optimization (PSO) and maximum likelihood (ML) based ones in terms of computational complexity. Simulations demonstrate that the proposed algorithm almost reaches the Bayesian Cramer-Rao bound (BCRB) and achieves a superior estimation accuracy by comparing to the PSO and the ML algorithms.
\end{abstract}


\IEEEpeerreviewmaketitle

\section{Introduction}
\subsection{Motivation and Literature Review}

The wireless communications is undergoing a significant transformation, marked by increased demands for wireless resources and adaptive intelligence. This shift is driven by the growing need for high-quality service and precise localization accuracy. Sectors like autonomous driving, smart transportation, and unmanned aerial vehicles (UAVs) exemplify this change, relying on attributes such as high data rates, unwavering reliability, and precise positioning. Meeting these demands requires the cultivation of innovative techniques to not only achieve precise localization but also facilitate high-speed communications. {\color{red}For instance, recent studies highlight the emergence of large antenna arrays as a transformative technology. In \cite{BayesianZhuTSP21}, Bayesian channel estimation techniques tailored for multi-user massive multiple input multiple output (MIMO) systems with extensive antenna arrays are explored. \cite{TSP22ZhangEstimation} introduces an innovative methodology for direction-of-arrival (DoA) estimation designed specifically for large antenna arrays, leveraging hybrid analog and digital architectures. This approach opens new avenues for optimizing spatial awareness in communication systems. Additionally, \cite{YangTWC21Communication} delves into communication and localization using extremely large lens antenna arrays.}Besides,  another promising technique is the use of reconfigurable intelligent surfaces (RIS), capable of altering the physical propagation environment to amalgamate signals at the receiver either destructively or constructively. The literature extensively reports on the applications of RIS in localization and communications.

The RIS is composed of numerous reflecting elements capable of actively modifying the phases and amplitudes of incident electromagnetic waves through a smart microcontroller, as highlighted in \cite{ElMossallamyReconfigurableTCMN20}. The cost-effectiveness of RIS hardware allows for its widespread use, providing additional controllable communication paths that enhance system performance in terms of reliability, energy/spectrum efficiency, and security, as discussed in \cite{GuoWeightedTWC20, ZhouSecureWCL21, LiIoTSecure21, TWCZhang22Maximization}. Consequently, RIS technology holds the promise of significantly improving wireless communications and localization, especially in the context of beyond fifth generation (B5G) or sixth generation (6G) communications, as emphasized in \cite{TVTRIS2022Bhowal, CMChen2022RIS}. A substantial body of research has been dedicated to exploring and harnessing the benefits and potentials of RIS-aided communications, reflecting the growing interest and recognition of its transformative impact \cite{RenzoJSAC20Smart, GacaninVTM20Toward}. {\color{red}
In recent literatures, a Bayesian framework was proposed in \cite{TengJSTSP22Bayesian} for user localization and tracking in RIS-aided MIMO systems. Delving into statistical methods for enhanced channel estimation accuracy, this work establishes a foundation for robust communication systems. Equally pivotal is the exploration in \cite{GanTC21Exploiting}, who delve into RIS-assisted multi-user multiple input single output (MISO) communications, emphasizing the exploitation of statistical channel state information (CSI) to optimize system performance. This article also draws insights from the study conducted in \cite{ZhaoyangTerahertz21JSAC} on multi-hop RIS-empowered terahertz communications, presenting a novel deep reinforcement learning based hybrid beamforming design and showcasing the versatility of RIS in the Terahertz frequency range. The comprehensive framework proposed in \cite{ProceedingLee22Chan} for channel estimation with RIS, further establishes the general applicability of RIS across diverse communication scenarios. Insights into robust channel estimation for RIS-aided millimeter-wave systems, addressing challenges such as RIS blockage, are contributed in \cite{RobustMaTVT22}. Additionally, \cite{WirelessTCP21} provides valuable perspectives on RIS-aided wireless communications, covering prototyping, adaptive beamforming, and real-world field trials.}

However, most of the prior works frequently assumed either perfect CSI or precise user location for RIS-aided systems that are obviously too optimistic for practical applications.  To address this issue, there exist a few works that are concerned with the imperfect CSI and inaccurate localization in RIS-aided systems. In what follows, the channel estimation and the user localization in RIS-aided systems are individually investigated.

With regard to the channel estimation of RIS-aided systems, the channels can be divided into far-field and near-field scenarios, as evidenced by recent studies. {\color{red}\cite{JiangHybrid12TWC} presents a pioneering study, introducing a hybrid far- and near-field modeling approach for reconfigurable intelligent surface (RIS) assisted Vehicle-to-Vehicle (V2V) channels. Their sub-array partition-based methodology emphasizes the intricate interplay between far-field and near-field effects, offering valuable insights for optimizing communication scenarios in V2V channels.  \cite{Cui23MIMOCM} provides a comprehensive exploration of near-field MIMO communications in the context of 6G evolution. \cite{PanJSTspRIS23} contributes to the field by focusing on RIS-aided near-field localization and channel estimation within the terahertz frequency range, which offers valuable insights for terahertz communication systems. \cite{GuerraTSPTracking21} explores near-field tracking with large antenna arrays, discussing fundamental limits and practical algorithms and contributes essential knowledge on the challenges and potential solutions associated with large antenna arrays in the context of near-field tracking applications.} Furthermore, channel estimation methods in the RIS-aided systems can be broadly categorized into parametric estimation methods and statistical estimation methods. In the domain of parametric channel estimation, various approaches address the sparsity or low-rank characteristics of RIS-aided system channels. Noteworthy works include \cite{ChannelHe21TWC, ChannelWei21CL, ArdahTRICESPL21, MatrixLiu20JSAC, CascadedLiu21WCL, ChannelWei21TC}, where methods such as message-passing algorithms, double-structured orthogonal matching pursuit (DS-OMP), and two-stage algorithms are proposed to estimate RIS-aided system channels while considering their inherent sparsity or low-rank properties. Additionally, works like \cite{JointZhou21TC, ChannelHe21TWC, CascadedLiu21WCL} present techniques involving alternating least squares, variational approximate message passing, atomic norm minimization, and wideband modeling to address RIS-aided channel estimation challenges. On the statistical front, similar efforts have been made to exploit RIS-aided communication systems. Examples include \cite{ChannelJSACYou20}, which estimates the instantaneous CSI of a single-user RIS-aided system using a hierarchical training reflection matrix design algorithm. In \cite{Joint22LiTC}, research delves into joint data detection and channel estimation for hybrid Reconfigurable Intelligent Surface (HRIS)-aided millimeter-wave orthogonal time-frequency space (OTFS) systems. Additionally, \cite{IntelligentCunhua22TWC} considers imperfect channel state information and correlated Rayleigh fading channels in the context of RIS-assisted multiple input single output (MISO) systems with hardware impairments. These statistical channel estimation approaches encompass a range of scenarios, offering insights into addressing challenges related to imperfect information and hardware impairments in RIS-assisted communication systems.

On the other hand, localization remains a critical concern in RIS-aided communication systems, with several studies shedding light on diverse aspects of this intricate challenge. Notably, \cite{WymeerschVTMRadio20} reported on far-field localization in both the uplink and downlink of RIS-aided systems. Further exploration in \cite{MaIndoorCL21} delved into indoor far-field localization scenarios, deriving the CRLB in closed-form and showcasing the RIS as a fundamental technique for achieving high indoor localization accuracy. \cite{GhatakOnCL21} analyzed the RIS-aided localization error bound, demonstrating superior performance compared to systems without RIS assistance. While some works, such as \cite{MetaLocalizationZhangTWC21} and \cite{ZhangTowardsCL21}, focused on localization algorithms for RIS-aided systems, channel estimation was neglected. Addressing this gap, Keykhosravi et al. \cite{KeykhosraviSISO21ICC} solved the synchronization and localization problem for RIS-aided single input single output (SISO) systems, assuming perfect CSI. They utilized maximum likelihood (ML) estimation by leveraging the dominance of the direct link signal power over the reflected signal power. ML-based estimation approaches were also proposed in \cite{ReconfigurableElzanatyTSP21}, accompanied by the derivation of corresponding CRLB. Moreover, RIS-aided localization challenges were explored in B5G \cite{WymeerschBeyond20ICC} and mm-Wave MIMO systems \cite{PositioningZhang20ICC}. Despite these endeavors, the localization of RIS-aided systems remains in its infancy, with numerous associated problems yet to be explored.

The majority of the aforementioned research works have predominantly concentrated on addressing either the channel estimation or localization challenges in the RIS-assisted communications. However, the intricacies arise as the joint estimation of channel states and localization becomes paramountly important, given the inherent coupling of user locations and channel estimation problems, which is a consequence of the shared environmental dependencies on channel gains, delays, and angles, significantly intensifying the complexity of the estimation problem. In a notable attempt to tackle this challenge, \cite{LinTWC22Channel} proposed a solution assuming a twin-RIS structure to facilitate channel estimation through tensor decomposition. The estimated CSI was then leveraged for the localization of far-field users. Similarly, in \cite{Han22STSPLocalization}, researchers delved into the intricacies of near-field joint localization and channel estimation, specifically in an extremely large RIS-aided MIMO system. Regrettably, these approaches were formulated under the assumption of specialized RIS structures, rendering them inapplicable to more general scenarios. The quest for effective methodologies that can accommodate diverse RIS configurations remains a pressing challenge in advancing the joint estimation of channel states and localization in RIS-assisted communications.

\subsection{Our Contributions}
In this paper, we dig into the complicated joint channel estimation and localization for RIS-aided wireless systems, focusing on a more general RIS-aided configuration. In this paper, the transmitter possesses only partial prior statistical knowledge regarding the user's location, channel gains, and the angle of departure (AoD). The challenge lies in the intricate interplay among the user's location, nonlinear phase shifts, channel gains, and AoD terms, rendering the joint estimation problem highly intricate. To address this complexity,
{\color{red} a variational Bayesian framework \cite{Arulampalam02TSPBayesian,RuanCCVector23}, which is also applied to the user detection and channel estimation \cite{ZhangTVT21Joint}, vehicle to vehicle channel estimation \cite{LiaoTITSMachine23}, signal recovery\cite{RajoriyaVariational23CL}, will be developed by capitalizing on the sparser angle pattern and the prior channel information.} In summary, the contributions of this paper are outlined as follows.
\begin{itemize}
  \item  Unlike \cite{Han22STSPLocalization} and \cite{LinTWC22Channel} that take into account the twin-RIS structure and extra large RIS requirements, a joint localization and channel estimation problem is considered for a general RIS-aided communication system with fewer constraints on the RIS structures.
  \item The sparser angle pattern and the prior channel information motivate us to develop a variational Bayesian learning-based framework of joint channel and location estimation. The proposed algorithm is applicable to both near-field but also far-field scenarios owing to the exploitation of the sparsity of the angular domain.
  \item
  The complexity analysis of the proposed algorithm is conducted in this paper. Since the joint channel and location estimation problem has a closed-form solution in each iteration, the proposed iterative algorithm converges faster than the PSO and ML-based ones.
 \item The BCRB of the joint estimation problem is derived to show the performance bounds of the joint estimation problem.
  Monte Carlo simulations demonstrate that the proposed algorithm almost reaches the Bayesian Cramer-Rao bound (BCRB).
\end{itemize}

\subsection{Organization}
The remainder of this paper is organized as follows. In Section II, the problem of joint channel estimation and localization in RIS-aided systems is formulated. A variational Bayesian learning-based joint channel and location estimation algorithm is proposed in Section III. 
Section IV carries out the complexity analysis of the proposed algorithm.
Finally, the simulation results are presented in Section V and the paper is concluded in Section VI.

\section{System Model}
We consider a RIS-aided system with an access point (AP) equipped with a single antenna and single antenna user in Fig.1. The RIS  is deployed for the aid of reflecting the signals from the AP to the user.
The position of the AP is ${\boldsymbol p}_{a} = \left[x_{a},y_{a},z_{a}\right]^T$ and the position of the user   is ${\boldsymbol p}_{u} = \left[x_{u},y_{u},z_{u}\right]^T$. We assume that the RIS is a uniform planar array with $M\times N$ reflecting elements. In the RIS, the inter-element distance between the column elements and the row elements are equal to $d$. The origin coordinate of the RIS is given by ${\boldsymbol p}_{r} = \left[x_{r},y_{r},z_{r}\right]^T$. The $\left(m,n\right)$-th element is located at ${\boldsymbol p}_{r}^{m,n} = {\left[ {{x_r} + \left( {m - 1} \right)d,{y_r},{z_r} + \left( {n - 1} \right)d} \right]^T}$ \. The user receives $L$ OFDM subcarriers both from the AP directly and from the RIS.  Similar to \cite{BoyuBayesian22STSP},  we assume the reflected paths always exist and the reflected signals are received by the user for localization and channel estimation.  Hence, the received signal at the user    side is given by \cite{STSPKamran22SISO, WymeerschBeyondICC20, KeykhosraviSISO21ICC},
\begin{equation}\label{receivedsignal}
\begin{aligned}
{\boldsymbol r}_t &= \underbrace {{\alpha _{au}}\sqrt {{P_w}} {\boldsymbol{\varphi }}\left( {{\zeta _{au}}} \right)}_{\color{red}{{\boldsymbol \Xi^t_{au}}}} \\
&+ \underbrace {{\alpha _{ru}}\sqrt {{P_w}} {\boldsymbol{\varphi }}\left( {{\zeta _{ru}}} \right){{\boldsymbol{a}}^T}\left( {\theta ,\vartheta } \right){\rm{diag}}\left( {{{\boldsymbol{\Omega }}_t}} \right){\boldsymbol{a}}\left( {\psi ,\phi } \right)}_{\boldsymbol \Xi _{ru}^t} + {{\boldsymbol{\varepsilon }}_t},
\end{aligned}
\end{equation}
where ${\varepsilon _t}$ is the zero-mean Gaussian noise with variance matrix $\delta {{\boldsymbol I}}$. ${\alpha _{au}}$ is the unknown complex channel gain of the AP-user link and ${\alpha _{ru}}$ is assumed to be an unknown complex channel gain of the AP-RIS-user link due to the random reflection in RIS.  $\sqrt{P_w}$ is the transmitted pilot symbol. {\color{red}${\boldsymbol \omega}_t = vec\left({\boldsymbol \Lambda}_t\right) \in {\mathbb C}^{MN \times 1}$ with the known phase shifts of the RIS elements at $t$-th transmission is given by ${\boldsymbol \Lambda}_t \in {\mathbb C}^{M \times N}$ and $\boldsymbol \Omega_t = \text{diag}\left({\boldsymbol \omega}_t\right)$}.
${\zeta _{au}}$ and ${\zeta _{ru}}$ are the delays of the AP-user   link and the AP-RIS-user   link respectively and respectively are given by
\begin{equation}\label{delayau}
{\zeta _{au}} = \frac{{{{\left\| {{{\boldsymbol p}_a} - {{\boldsymbol p}_u}} \right\|}_2}}}{c},
\end{equation}
and
\begin{equation}\label{delaru}
{\zeta _{ru}} = \frac{{{{\left\| {{{\boldsymbol p}_a} - {{\boldsymbol p}_r}} \right\|}_2}}+ {{\left\| {{{\boldsymbol p}_r} - {{\boldsymbol p}_u}} \right\|}_2}}{c},
\end{equation}
and the phase shifts caused by the delays are given by ${\boldsymbol{\varphi }}\left( {{\zeta _{au}}} \right) = {\left[ {1,{e^{ - j2\pi {\zeta _{au}}{\Delta f}}} \cdots,{e^{ - j2\pi \left( {L - 1} \right){\zeta _{au}}{\Delta f}}}} \right]^T}$ and ${\boldsymbol{\varphi }}\left( {{\zeta _{ru}}} \right) = {\left[ {1,{e^{ - j2\pi {\zeta _{ru}}{\Delta f}}} \cdots,{e^{ - j2\pi \left( {L - 1} \right){\zeta _{ru}}{\Delta f}}}} \right]^T}$ respectively. ${\Delta f}$ is the frequency spacing. In \eqref{receivedsignal}, the steering vector ${{\boldsymbol a}}\left( {\theta ,\vartheta } \right) \in {\mathbb C}^{MN \times 1}$ is given by
\begin{equation}\label{steeringvector1}
{\boldsymbol{a}}\left( {\theta ,\vartheta } \right) = {{\boldsymbol{a}}_r}\left( \theta,\vartheta  \right) \otimes {{\boldsymbol{a}}_c}\left( \theta,\vartheta  \right),
\end{equation}
where $\otimes$ is the Kronecker product and $\theta,\vartheta$ are the azimuth and elevation angles from the AP to the RIS and these angles are assumed to be known. The  ${{\boldsymbol{a}}_r}\left( \theta,\vartheta  \right)$ and ${{\boldsymbol{a}}_c}\left( \theta,\vartheta  \right)$ are respectively given by
\begin{equation}\label{wavevector1}
 {{\boldsymbol{a}}_r}\left( {\theta ,\vartheta } \right) = \left[ {1,{e^{  j\tau }}, \cdots ,{e^{ j\left( {M - 1} \right)\tau }}} \right]^T,
\end{equation}
and
\begin{equation}\label{wavevector2}
{{\boldsymbol{a}}_c}\left( {\theta ,\vartheta } \right) = \left[ {1,{e^{  j\nu}}, \cdots ,{e^{ j\left( {N - 1} \right)\nu}}} \right]^T,
\end{equation}
where $\tau = \frac{2\pi d}{\lambda}\cos \vartheta$ and $\nu = \frac{2\pi d}{\lambda} \cos \theta \sin \vartheta$.

Similarly ${{\boldsymbol a}}\left( {\psi ,\phi } \right)\in {\mathbb C}^{MN \times 1}$  can be given by
\begin{equation}\label{steeringvector1}
{\boldsymbol a}\left( {\psi ,\phi } \right) = {{\boldsymbol{a}}_r}\left({\psi ,\phi }  \right) \otimes {{\boldsymbol{a}}_c}\left({\psi ,\phi } \right).
\end{equation}
with
\begin{equation}\label{wavevector3}
 {{\boldsymbol{a}}_r}\left( {\psi ,\phi } \right) = \left[ {1,{e^{  j\upsilon }}, \cdots ,{e^{ j\left( {M - 1} \right)\upsilon }}} \right]^T,
\end{equation}
and
\begin{equation}\label{wavevector4}
{{\boldsymbol{a}}_c}\left( {\psi ,\phi } \right) = \left[ {1,{e^{  j\kappa}}, \cdots ,{e^{  j\left( {N - 1} \right)\kappa}}} \right]^T.
\end{equation}
where $\upsilon = \frac{2\pi d}{\lambda}\cos \psi  $ and $\kappa = \frac{2\pi d}{\lambda} \cos \psi\sin \phi$ for the far-field scenarios and
\begin{equation}\label{nearfiled}
 {\boldsymbol{a}}\left( {\psi ,\phi } \right) = \exp \left\{ {j\frac{{2\pi }}{\lambda }\left[ {{\boldsymbol \digamma ^T}{\boldsymbol{p}}_r^{m,n} - \frac{1}{{2{d_{m,n}}}}{{\tilde {\boldsymbol \digamma}}^T}{\boldsymbol{\tilde p}}_r^{m,n}} \right]} \right\},
\end{equation}
for the near-field scenarios\footnote{It means the RIS and the user are in far-field or near-field scenarios. The steering vector for near field scenarios is an approximation to the exact one according to \cite{Liu14MatrixSJ}. {\color{red}For channels encompassing mixed near and far-field components, we extend the system model to accommodate multiple users, as outlined in \cite{TC23LUHierarchical}. However, it's worth noting that the multiple users scenario necessitates the phase optimization of RIS elements, a task that exceeds the feasibility scope of the proposed algorithm.}} and ${{d_{m,n}}}$ is the distance between the $\left(m,n\right)$-th element and the origin coordinate of RIS.  ${\boldsymbol{\tilde p}}_r^{m,n} = {\left[ {{\boldsymbol{p}}_r^{m,n} - {{\boldsymbol{p}}_r}} \right]^2}$ and ${\left[ \cdot \right]^2}$ represents the elementwise square.  ${\boldsymbol{\digamma}} = {\left[ {\cos \phi \sin \psi ,\sin \phi \sin \psi ,\cos \psi } \right]}$  and
$\widetilde {\boldsymbol{\digamma}} = {\left[ {1 - {{\cos }^2} \phi {{\sin }^2} \psi,1 - {{\sin }^2}\phi{{\sin }^2}\psi, {{\sin }^2}\psi} \right]}$.  The angles $\phi$ and $\psi$ are assumed to be unknown in both scenarios.

By collecting the $T$ snapshots of the received signal ${\boldsymbol R} = \left[{\boldsymbol r}_1,...,{\boldsymbol r}_T\right]$, the system model can be given by
\begin{equation}\label{newsystem}
{\boldsymbol{R}} = \underbrace {{\alpha _{au}}\sqrt {{P_w}} {\boldsymbol{\varphi }}\left( {{\zeta _{au}}} \right){{\boldsymbol{1}}^T}}_{{{\boldsymbol{\Xi }}_{au}}} + \underbrace {{\alpha _{ru}}\sqrt {{P_w}} {\boldsymbol{\varphi }}\left( {{\zeta _{ru}}} \right){{\left( {{\boldsymbol{\Upsilon a}}\left( {\psi ,\phi } \right)} \right)}^T}}_{{{\boldsymbol{\Xi }}_{ru}}} + {\boldsymbol{\varepsilon }},
\end{equation}
where ${\boldsymbol{\Upsilon }} = {\left[ {{\boldsymbol{\Upsilon }}_1^T, \cdots ,{\boldsymbol{\Upsilon }}_{T}^T} \right]^T}$ and $\boldsymbol \Upsilon_t  = {{\boldsymbol{a}}^T}\left( {\theta ,\vartheta } \right){{diag}}\left( {\boldsymbol{\Omega }_t} \right)$.  The likelihood function can be given by
\begin{equation}\label{likelihood}
\begin{aligned}
& p\left( {{\boldsymbol{R}}|\boldsymbol \Theta } \right) = \prod\limits_t^T {p\left( {{\boldsymbol{r}}\left( t \right)|\boldsymbol \Theta } \right)} \\
& \propto \prod\limits_{t = 1}^T {\exp \left( { - \frac{1}{{2\delta }}{{\left( {{\boldsymbol{r}}_t - {\color{red}{{\boldsymbol \Xi^t_{au}}}} - {\boldsymbol{\Xi }}_{ru}^t} \right)}^H}\left( {{\boldsymbol{r}}_t - {\color{red}{{\boldsymbol \Xi^t_{au}}}} - {\boldsymbol{\Xi }}_{ru}^t} \right)} \right)},
\end{aligned}
\end{equation}
where $\boldsymbol \Theta  = {\left[ { \boldsymbol \varphi ^T\left( {{\zeta _{au}}} \right),\boldsymbol \varphi^T \left( {{\zeta _{ru}}} \right),\psi ,\phi, \alpha_{au},{\alpha _{ru}} } \right]}$.
In the paper, we focus on the estimation of the user location and channel state information with the aid of the RIS. In \eqref{likelihood}, the direct maximization is intractable due to two extremely challenging problems: the coupling unknown parameter and the nonlinear steering vector ${\boldsymbol a}\left( {\psi ,\phi } \right)$.  To obtain the solution of the user location and CSI,  we proposed a variational Bayesian inference-based estimation algorithm.
\begin{figure}
  \centering
  \includegraphics[width=3.0in]{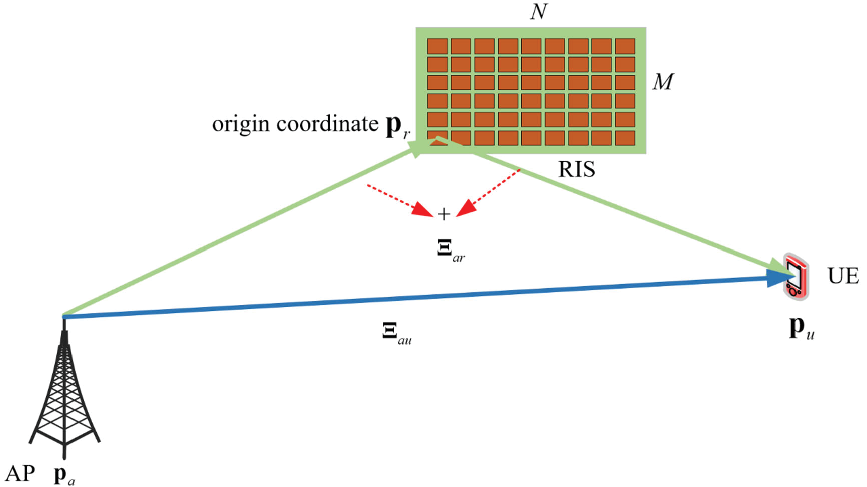}\\
  \caption{The system model}\label{1}
\end{figure}

\section{Variational Bayesian Learning-based Localization and Channel Estimation Algorithm}
\subsection{Sparse Representation}
Considering the sparsity of the angles in ${\boldsymbol a}\left( {\psi ,\phi } \right)$,  the system model in \eqref{steeringvector1} is reformulated via sparse representation.
First, the angle spread of ${\psi ,\phi }$ are both assumed to be $\left[-\frac{\pi}{2},\frac{\pi}{2}\right]$ and the spread can be both equally divided into $P$ and $Q$ resolutions respectively,  then we can obtain
\begin{equation}\label{sparseGamma}
{\boldsymbol \Gamma} = \begin{bmatrix}
\left(\bar \psi_{1}, \bar \phi_{1}\right )&  \cdots& \left(\bar \psi_{1}, \bar \phi_{Q}\right )\\
 \vdots &  & \\
 \left(\bar \psi_{P}, \bar \phi_{1}\right )& \cdots & \left(\bar \psi_{P}, \bar \phi_{Q}\right )
\end{bmatrix}\in {\mathbb{C}}^{P\times Q}.
\end{equation}

Using the vectorization of \eqref{sparseGamma}, it yields
\begin{equation}\label{vectorGamma}
\bar{\boldsymbol \theta} = \begin{bmatrix}
\left(\bar \psi_{1}, \bar \phi_{1}\right ) & \cdots & \left(\bar \psi_{1}, \bar \phi_{Q}\right ) & \cdots & \left(\bar \psi_{P}, \bar \phi_{Q}\right )
\end{bmatrix}\in {\mathbb{C}}^{1 \times PQ}.
\end{equation}

Embedding \eqref{vectorGamma} into \eqref{steeringvector1}, we can obtain
\begin{equation}\label{re_steeringvector1}
\bar{\boldsymbol A} = {\boldsymbol{a}}\left(\bar {\boldsymbol  \theta} \right) =  \left[ {\begin{array}{*{20}{c}}
{{\boldsymbol{a}}\left( {{{\bar \psi }_1},{{\bar \phi }_1}} \right)}& \cdots &{{\boldsymbol{a}}\left( {{{\bar \psi }_P},{{\bar \phi }_Q}} \right)}
\end{array}} \right]\in {\mathbb{C}}^{MN \times PQ}.
\end{equation}

Though the true angles are continuous variables and may not fall on the grid points, the off-grid errors can be ignorable given enough resolutions of $P$ and $Q$.  Hence, we apply the on-grid model and the steering vector is given by
\begin{equation}\label{discretization}
\begin{aligned}
\boldsymbol{\cal {A}}& =  \left[ {\begin{array}{*{20}{c}}
{{\boldsymbol{a}}\left( {{{\bar \psi }_1},{{\bar \phi }_1}} \right)}& \cdots, {{\boldsymbol{a}}\left( {{{\bar \psi }_p},{{\bar \phi }_q}} \right)}, &\cdots {{\boldsymbol{a}}\left( {{{\bar \psi }_P},{{\bar \phi }_Q}} \right)}
\end{array}} \right]\\ &\in {\mathbb{C}}^{MN \times PQ}.
\end{aligned}
\end{equation}

Therefore, the reflected link $\Xi^t_{ru}$ can be reformulated as
\begin{equation}\label{re_steering}
{{\boldsymbol \Xi}^t_{ru}} =  {\sqrt {P_w}}{\boldsymbol \varphi} \left( {{{ \zeta} _{ru}}} \right)\boldsymbol \Upsilon_t \boldsymbol{\cal {A}} {\boldsymbol \Delta _{ru}},
\end{equation}
where ${{\boldsymbol \Delta}_{ru}}\in {\mathbb C}^{PQ \times 1}$ is a vector with only one unknown non-zero element ${\alpha _{ru}}$ at unknown location of ${{\boldsymbol \Delta} _{ru}}$. The system model can be approximated as
\begin{equation}\label{new1model}
{\boldsymbol{R}} = \underbrace {{\alpha _{au}}{{\sqrt {P_w} }}\boldsymbol \varphi \left( {{\zeta _{au}}} \right){{\boldsymbol 1}^T}}_{{{\boldsymbol{\Xi }}_{au}}} +\underbrace {\sqrt {{P_w}} {\boldsymbol{\varphi }}\left( {{\zeta _{ru}}} \right){{\left( {{\boldsymbol{\Upsilon {\cal {A}}}}{{\boldsymbol{\Delta }}_{ru}}} \right)}^T}}_{{{\boldsymbol{\Xi }}_{ru}}} + {\boldsymbol{\varepsilon }},
\end{equation}
where ${{\boldsymbol 1}}\in \mathbb{R}^{T \times 1}$ is a column vector and ${\boldsymbol{\varepsilon }} = {\left[ {{\bf{\varepsilon }}_1^T, \cdots ,{\bf{\varepsilon }}_{T }^T} \right]^T}$.
Hence, the likelihood function of \eqref{new1model} can be given by
\begin{equation}\label{likelihoodnew}
\begin{aligned}
&p\left( {{\boldsymbol{R}}|{\boldsymbol \nu}} \right) \\ & \propto \exp \left( { - {{\left( {{\boldsymbol{R}} - {{\boldsymbol{\Xi }}_{au}} - {{\boldsymbol{\Xi }}_{ru}}} \right)}^H}{{\boldsymbol \Sigma}^{ - 1}}\left( {{\boldsymbol{R}} - {{\boldsymbol{\Xi }}_{au}} - {{\boldsymbol{\Xi }}_{ru}}} \right)} \right),
\end{aligned}
\end{equation}
where  ${\boldsymbol \nu}  = \left[ {\boldsymbol \varphi ^T\left( {{\zeta _{au}}} \right),\boldsymbol \varphi ^T\left( {{\zeta _{ru}}} \right),\psi ,\phi ,{\alpha_{au}}, {{\boldsymbol \Delta}_{ru}}}\right]$ and ${\boldsymbol \Sigma}$ is the diagonal covariance matrix with all diagonal elements $\delta$.

The UE location estimation is via the maximum likelihood estimation in \eqref{likelihoodnew} and the main objective is to estimate the parameters $\phi$, $\psi$, $\zeta_{au}$, $\zeta_{ru}$.  For the angles $\psi,\phi$, they are represented sparsely in \eqref{re_steering} and it is equivalent to estimate the nonzero elements variable in the sparse vector ${{\boldsymbol \Delta}_{ru}}$. For the delays $\zeta_{au}$ and $\zeta_{ru}$, it is intractable to acquire the closed-form solution via direct maximization of \eqref{likelihoodnew}.  Considering the estimation of unknown parameters via maximum a posterior (MAP), the prior distributions of the unknown parameters require further clarifications:
\begin{itemize}
  \item The line-of-sight complex channel gain $\alpha_{au}$ is assumed to subject to a complex Gaussian distribution as
  \begin{equation}\label{prior_distributionofPsis}
p\left( {{\alpha _{au}}} \right) = \mathcal{CN}\left( {{\alpha _{au}}|{\mu _{{\alpha _{au}}}},{\delta _{{\alpha _{au}}}}} \right).
  \end{equation}
  \item In the paper, the line-of-sight time delay ${\zeta _{au}}$ and the reflecting path time delay ${\zeta _{ru}}$  are nonlinear to the likelihood function \eqref{likelihood} and it is difficult to find the closed-form estimation to the time delays. Hence, we turn to estimate the two phase shifts $\boldsymbol \varphi \left( {{\zeta _{au}}} \right)$ and $\boldsymbol \varphi \left( {{\zeta _{ru}}} \right)$. However, it is challenging to obtain the precise prior distributions of the phase shift variables.
       Hence, we assume the non-informative complex Gaussian distributions
      \begin{equation}\label{prior_distributionofPsit}
      p\left( {{{\boldsymbol{\Psi }}_i}} \right) = \mathcal{CN}\left( {{{{\boldsymbol \Psi }}_i}|{\boldsymbol \mu _{{{{\boldsymbol \Psi }}_i}}},{\boldsymbol \Sigma _{{{{\boldsymbol \Psi }}_i}}}} \right),
      \end{equation}
 with ${\delta _{{{{\boldsymbol \Psi }}_i}}}  \rightarrow \infty$  and ${{\boldsymbol{\Psi }}_i} \in \left[ {\boldsymbol \varphi \left( {{\zeta _{au}}} \right),\boldsymbol \varphi \left( {{\zeta _{ru}}} \right)} \right]$. In practice, the variance can be replaced by relatively large positive values.
  \item For the sparse representation parameter ${{\boldsymbol \Delta}_{ru}}$, only one nonzero element exists at an unknown location and the other elements are zeros. The nonzero element is also unknown. Therefore, it is  assumed   that each element ${{\boldsymbol \Delta}^{i}_{ru}}$  in the  sparse vector ${{\boldsymbol \Delta}_{ru}}$ follows a mixture Gaussian  distribution \cite{SparseChang21STSP} as follows
       \begin{equation}\label{prior_Delta}
p\left( {{{\boldsymbol{\Delta }}_{ru}}} \right) = \prod\limits_{i = 1}^{{\rm{PQ}}} p \left( {{\boldsymbol{\Delta }}_{ru}^i} \right) = \prod\limits_{i = 1}^{{\rm{PQ}}} {\prod\limits_{l = 1}^2 {{\cal C}{\cal N}{{\left( {{\boldsymbol{\Delta }}_{ru}^i|\mu _{\boldsymbol{\Delta }}^l,w_{{\boldsymbol{\Delta }}_{ru}^i}^{ - 1}} \right)}^{{g_{i,l}}}}} }
      \end{equation}
      where complex Gaussian distribution with $\mu _{\boldsymbol{\Delta }}^1= 0$ and $w_{{\boldsymbol{\Delta }}_{ru}^i}^{ - 1} \gg \mu _{\boldsymbol{\Delta }}^1$ is to enforce a prior distribution to the zero elements. ${{\boldsymbol g}_i} = \left[ {{g_{i,1}},{g_{i,2}}} \right]$ is indicator vector and is given by
      \begin{equation}\label{indicator}
      {{\boldsymbol g}_i} = \left\{ {\begin{array}{*{20}{c}}
{(1,0)}&{\boldsymbol \Delta _{ru}^i \ne 0,}\\
{(0,1)}&{\boldsymbol \Delta _{ru}^i = 0.}
\end{array}} \right.
      \end{equation}
\item The inverse variance ${{w _{{\boldsymbol{\Delta }}_{ru}^i}}}$in \eqref{prior_Delta}  is further constrained by imposing a prior distribution and we assume a inverse variance ${{w _{{\boldsymbol{\Delta }}_{ru}^i}}}$, which is given by
 \begin{equation}\label{Gamma_prior}
 p\left( {w}_{\boldsymbol \Delta }\right) =  \prod\limits_{i = 1}^{\text{PQ}} p\left( {{w _{{\boldsymbol{\Delta }}_{ru}^i}}} \right) =  \prod\limits_{i = 1}^{\text{PQ}} \Gamma  \left( {{w_{{\boldsymbol{\Delta }}_{ru}^i}}|a_i,b_i} \right),
 \end{equation}
where $\Gamma\left(\cdot\right)$ is the Gamma distribution and $a_i$, $b_i$ are the known parameters of the Gamma distribution and ${w}_{\boldsymbol \Delta }  = \left( {{w _{{\boldsymbol{\Delta }}_{ru}^1}}, \cdots ,{w _{{\boldsymbol{\Delta }}_{ru}^{\text{PQ}}}}} \right)$.

 \item Therefore, the indicator variable ${\boldsymbol g}_{i}$ can be modeled to follow a non-informative  categorical distribution, which is given by
\begin{equation}\label{multidis}
p\left( {{{\boldsymbol r}}|{\boldsymbol \chi }} \right) = \prod\limits_{i = 1}^{\text{PQ}}\prod\limits_{l = 1}^{2} p\left( {{{g}_{i,l}}|{\boldsymbol \chi }_l} \right) = \prod\limits_{i = 1}^{\text{PQ}}\prod\limits_{l = 1}^{2} {\chi _{l}^{{g_{i,l}}}},
\end{equation}
with ${\boldsymbol g} = \left[{\boldsymbol g}_1,\cdots,{\boldsymbol g}_{\text{PQ}}\right]$ and ${{\boldsymbol{\chi }}} = \left[ {{\chi _{1}},{\chi _{2}}} \right] = \left[ {\frac{1}{{\text{PQ}}}},1 - \frac{1}{{\text{PQ}}} \right]$. For easy presentation, we denote an unknown variable vector {${\boldsymbol {\cal W}}  = \left[ {\boldsymbol \varphi ^T\left( {{\zeta _{au}}} \right),\boldsymbol \varphi^T \left( {{\zeta _{ru}}} \right),\psi ,\phi ,{{\boldsymbol \Delta}_{ru}}, \boldsymbol g, {w}_{\boldsymbol \Delta }}\right]$.}
\end{itemize}

\subsection{Variational Bayesian Learning Framework}
Based on the problem reformulation, our goal is to learn the true posterior distribution of the channel parameters and locations. Then the delay parameters, the online angles and the channel gains can be estimated as a maximum posterior problem as follows
\begin{equation}\label{map}
\boldsymbol {\hat {\cal W}} = \text{argmax}\int {p\left( {{\boldsymbol{R}}|\boldsymbol {\cal W}} \right)p\left( \boldsymbol {\cal W} \right)} d{\boldsymbol{g}}d{w_{\boldsymbol{\Delta }}},
\end{equation}
where involves numerous prior distributions, multiple integrals, coupled channel and location parameters, and the nonlinear non-convex objective function. Thus it is intractable to directly obtain learning features and find a closed-form solution. Hence, we focus on finding an approximation distribution to the true posterior distribution, which is also tractable for the MAP or MMSE estimators.

In the variational Bayesian learning framework, we aim at finding a variational $q( \boldsymbol {\cal W} )$ to the posterior distribution $p( \boldsymbol {\cal W} |{\boldsymbol{R}})$ in (\ref{map}) and the variational distribution $q( \boldsymbol {\cal W} )$ is tractable.
Revoked by the mean-field theory and assumption, we factorize the variational distribution $q( \boldsymbol {\cal W} )$ as
\begin{equation}\label{meanfield}
q\left(  \boldsymbol {\cal W}  \right) = \prod\limits_{{ {\boldsymbol {\cal W}}_k} \in  \boldsymbol {\cal W} } {q\left( {{{\boldsymbol {\cal W}} _k}} \right)} ,
\end{equation}
To measure the approximation between the variational distribution $q({\boldsymbol {\cal W}})$ and the true distribution $p({\boldsymbol {\cal W}}|{\boldsymbol{R}})$,
Kullback-Leibler (KL) divergence \cite{fox2012tutorial} is introduced and minimized as
\begin{equation}\label{KL}
\begin{aligned}
{\text{KL}}\left( {q\left( {\boldsymbol {\cal W}} \right)||p\left( {\boldsymbol {\cal W}} | {\boldsymbol R}\right)} \right)
 =  - {\mathbb{E}_{q\left( {\boldsymbol {\cal W}} \right)}}\left\{ {\ln \frac{{p\left({\boldsymbol {\cal W}} | {\boldsymbol R}\right)}}{{q\left({\boldsymbol {\cal W}} \right)}}} \right\} \ge 0,
 \end{aligned}
\end{equation}
where ${\mathbb{E}_{q\left({\boldsymbol {\cal W}} \right)}}$ is the expectation with respect to ${q\left({\boldsymbol {\cal W}} \right)}$ and the equality holds only when $q\left({\boldsymbol {\cal W}} \right) = p\left(  {\boldsymbol {\cal W}} | {\boldsymbol R} \right)$.
Based on the mean-field theory in \eqref{meanfield} and the alternative optimization method, the variational distribution can be iteratively approximated as \cite{TCRobustLi2020}
\begin{equation}\label{variationalphik}
q^{(\xi)}\left( {{{\boldsymbol {\cal W}} _k}} \right) \propto \exp \left\{ {{\mathbb{E}_{{q^{(\xi)}\left( {{{\boldsymbol {\cal W}} _{\backslash k}}} \right)} }}\left[ {\ln p\left( {{\boldsymbol {\cal W}} ,{\boldsymbol{R}}} \right)} \right]} \right\},
\end{equation}
where $q^{(\xi)}\left( {{{\boldsymbol {\cal W}} _k}} \right)$ is the approximation in the $\xi$-th iteration and $p\left( {{\boldsymbol {\cal W}} ,{\boldsymbol{R}}} \right)$ is the joint probability. ${\mathbb{E}_{{q^{(\xi )}}\left( {{{\boldsymbol {\cal W}}_{\backslash k}}} \right)}}$ means the expectation with respect the variational distributions excluding the variational distribution $q^{(\xi)}\left( {{{\boldsymbol {\cal W}} _k}} \right)$.
The approximated distribution ${q\left( {{{\boldsymbol {\cal W}}_k}} \right)}$ in fact can be regarded as the approximation of the corresponding posterior distribution $p\left(  {\boldsymbol {\cal W}}_k | {\boldsymbol{R}} \right)$. For example, $q\left({{{\boldsymbol g}}} \right) $ is the approximation to the posterior distribution $p\left( { {{{{\boldsymbol g}}}}|{\boldsymbol{R}}} \right) $. Then the MAP estimation of each parameter ${{\boldsymbol {\cal W}}_k}$ can be achieved as
\begin{equation}\label{MAP_estimation}
{{\boldsymbol {\cal W}}^{{\text{MAP}}}_k}= {\mathop{\rm argmax}\nolimits} \; {q\left( {{{\boldsymbol {\cal W}}_k}} \right)}.
\end{equation}

{To learn the tractable forms of variational distribution $q\left( {{{\boldsymbol {\cal W}}}} \right)$, we assume the prior distributions and the variational distribution follows the conjugate prior principles, which renders the variational distribution $q^{(\xi)}\left( {{{\boldsymbol {\cal W}} _k}} \right)$ is identical to the prior distribution $p\left( {{{\boldsymbol {\cal W}} _k}} \right)$ in form.}

The proposed Bayesian framework can learn the true posterior distribution via the alternative optimization of the KL-divergence. Given the learning distribution, the channel parameters and localization can be done via posterior estimators. In the following subsections, the detailed variational distributions are derived and the user location is estimated iteratively via the estimation of other parameters.

\subsection{Estimation of Channel Gains ${\alpha _{au}}$ }\label{es_alpha}
First, we consider the estimation problem of LOS channel gain ${\alpha _{au}}$. According to \eqref{variationalphik}, the $\xi$-th iteration variational distribution $q^{(\xi)}\left( {{{\alpha} _{au}}} \right)$ can be given by
\begin{equation}\label{variaotionalalpha_au}
\begin{aligned}
{q^{\left( \xi  \right)}}\left( {{\alpha _{au}}} \right)  \propto \exp \left\{ {\mathbb{E}_{ {{q^{(\xi )}}\left( {{{\boldsymbol {\cal W}}_{\backslash {{\alpha} _{au}}}}} \right)} }}\left( {\ln p\left( {{\boldsymbol{R}}|{\boldsymbol {\cal W}}} \right)} \right) \right.
\\
\phantom{=\;\;}
\left.+  {\mathbb{E}_{{{q^{(\xi )}}\left( {{{\boldsymbol {\cal W}}_{\backslash {{\alpha} _{au}}}}} \right)} }}\left( {\ln p\left( {{\alpha _{au}}} \right)} \right) \right\}.
\end{aligned}
\end{equation}

By plugging likelihood function in \eqref{likelihoodnew}, the first variational expectation can be given by
\begin{equation}\label{expectation_alphaau1}
\begin{aligned}
&{\mathbb{E}_{{q^{(\xi )}}\left( {{{\boldsymbol{W}}_{\backslash {\alpha _{au}}}}} \right)}}\left( {\ln p\left( {{\boldsymbol{R}}|{\boldsymbol{W}}} \right)} \right) \\ &= \sum\limits_{t = 1}^T {{\mathbb{E}_{{q^{(\xi )}}\left( {{{\boldsymbol{W}}_{\backslash {\alpha _{au}}}}} \right)}}} tr\left( { - \frac{{{\boldsymbol{\Xi }}_{au}^H{{\boldsymbol{\Xi }}_{au}} - 2{{\left( {{{\boldsymbol{r}}_t} - {\boldsymbol{\Xi }}_{ru}^t} \right)}^H}{{\boldsymbol{\Xi }}_{au}}}}{{2\delta }}} \right) \\&+ {\cal C},
\end{aligned}
\end{equation}
where $\cal C$ are the terms that can be regarded as constant and
\begin{equation}\label{expectation_alphaau2}
\begin{aligned}
{\mathbb{E}_{{q^{(\xi )}}\left( {{{\boldsymbol {\cal W}}_{\backslash {\alpha _{au}}}}} \right)}}\left( {{\boldsymbol{\Xi }}_{au}^H{{\boldsymbol{\Xi }}_{au}}} \right)
=\alpha _{au}^H\alpha _{au}{P_w}\mathcal{B}_{{{\zeta}_{au}}}^{\left( {\xi } \right)},
\end{aligned}
\end{equation}
where the scalar $\mathcal{B}_{{{\zeta}_{au}}}^{\left( {\xi } \right)}= {\mathbb{E}_{{q^{(\xi )}}\left( {{{\boldsymbol {\cal W}}_{\backslash {{{\alpha }}_{au}}}}} \right)}}\left( {{{\boldsymbol{\varphi }}^H}\left( {{\zeta _{au}}} \right){\boldsymbol{\varphi }}\left( {{\zeta _{au}}} \right)} \right)$ involves the expectation with respect to the nonlinear delay terms $\varphi \left( {{\zeta _{au}}} \right)$ and the parameter $\mathcal{B}_{{{\zeta}_{au}}}^{\left( {\xi } \right)}$ can be given by
\begin{equation}\label{Bexpress}
\begin{aligned}
{\cal B}_{{\zeta _{au}}}^{\left( \xi  \right)} &= {\mathbb{E}_{{q^{(\xi )}}\left( {{{\boldsymbol {\cal W}}_{\backslash {\alpha _{au}}}}} \right)}}\left( {{{\boldsymbol{\varphi }}^H}\left( {{\zeta _{au}}} \right){\boldsymbol{\varphi }}\left( {{\zeta _{au}}} \right)} \right) \\ &= {\left( {{\boldsymbol{\mu }}_{{\boldsymbol{\varphi }}\left( {{\zeta _{au}}} \right)}^{\left( \xi  \right)}} \right)^H}{\boldsymbol{\mu }}_{{\boldsymbol{\varphi }}\left( {{\zeta _{au}}} \right)}^{\left( \xi  \right)} + tr\left( {{\boldsymbol{\Sigma }}_{{\boldsymbol{\varphi }}\left( {{\zeta _{au}}} \right)}^{\left( \xi  \right)}} \right),
\end{aligned}
\end{equation}
where ${\boldsymbol{\mu }}_{{\boldsymbol{\varphi }}\left( {{\zeta _{au}}} \right)}^{\left( \xi  \right)}$ and ${{\boldsymbol{\Sigma }}_{{\boldsymbol{\varphi }}\left( {{\zeta _{au}}} \right)}^{\left( \xi  \right)}}$ are the mean and variance of $\xi$-th distribution ${q^{\left( \xi  \right)}}\left( {{\boldsymbol{\varphi }}\left( {{\zeta _{au}}} \right)} \right)$ and
\begin{equation}\label{variational_alphaau}
{q^{\left( \xi  \right)}}\left( {{\boldsymbol{\varphi }}\left( {{\zeta _{au}}} \right)} \right) = \mathcal{CN}\left( {{\boldsymbol{\varphi }}\left( {{\zeta _{au}}} \right)|{\boldsymbol {\mu }}_{{\boldsymbol{\varphi }}\left( {{\zeta _{au}}} \right)}^{\left( \xi  \right)},{\boldsymbol{\Sigma }}_{{\boldsymbol{\varphi }}\left( {{\zeta _{au}}} \right)}^{\left( \xi  \right)}} \right).
\end{equation}

In \eqref{expectation_alphaau1}, the expectation term ${\mathbb{E}_{{q^{(\xi )}}\left( {{{\boldsymbol {\cal W}}_{\backslash {\alpha _{au}}}}} \right)}}\left( {{{\boldsymbol{\Xi }}^t_{au}}} \right)$ also involves the phase shift ${{\boldsymbol{\varphi }}\left( {{\zeta _{ru}}} \right)}$.
Thus the expectation term ${\mathbb{E}_{{q^{(\xi )}}\left( {{{\boldsymbol {\cal W}}_{\backslash {\alpha _{au}}}}} \right)}}\left( {{{\boldsymbol{\Xi }}_{au}}} \right) $ is  given by
\begin{equation}\label{expectation_alphaau3}
\begin{aligned}
{\mathbb{E}_{{q^{(\xi )}}\left( {{{\boldsymbol {\cal W}}_{\backslash {\alpha _{au}}}}} \right)}}\left( {{{\boldsymbol{\Xi }}_{au}}} \right) = {\alpha _{au}}\sqrt {{P_w}} {\boldsymbol{\mu }}_{{\boldsymbol{\varphi }}\left( {{\zeta _{au}}} \right)}^{\left( \xi  \right)},
\end{aligned}
\end{equation}

\begin{equation}\label{expectation_alphaau4}
\begin{aligned}
{\mathbb{E}_{{q^{(\xi )}}\left( {{{\boldsymbol {\cal W}} _{\backslash {\alpha _{au}}}}} \right)}}\left( {{\boldsymbol{r}}_t - {\boldsymbol{\Xi }}_{ru}^t} \right) = {\boldsymbol{r}}_t - \underbrace {\sqrt {{P_w}} {\boldsymbol{\mu }}_{{\boldsymbol{\varphi }}\left( {{\zeta _{ru}}} \right)}^{\left( \xi  \right)}{{\boldsymbol{\Upsilon }}_t}{\cal A}{\boldsymbol{\mu }}_{{{\boldsymbol{\Delta }}_{ru}}}^{\left( \xi  \right)}}_{{\boldsymbol{\Theta }}_{{\alpha _{ru}}}^{\left( {t,\xi } \right)}},
\end{aligned}
\end{equation}
where ${\boldsymbol \mu}_{{{\boldsymbol \Delta}_{ru}}}^{\left( \xi  \right)}$ is  the mean of the variational distribution ${q^{\left( \xi  \right)}}\left( {{\boldsymbol{\Delta }}_{ru}} \right)$ and  is given by
\begin{equation}\label{variational_disDel}
\begin{aligned}
{q^{\left( \xi  \right)}}\left( {{\boldsymbol{\Delta }}_{ru}} \right) &\propto {\cal C}{\cal N}\left( {{\boldsymbol{\Delta }}_{ru}|\boldsymbol \mu _{{\boldsymbol{\Delta }}_{ru}}^{\left( \xi  \right)},\boldsymbol \Sigma _{{\boldsymbol{\Delta }}_{ru}}^{\left( \xi  \right)}} \right)  \\&= \prod\limits_{i=1}^{\text{PQ}}{\cal C}{\cal N}\left( {{\boldsymbol{\Delta }}^i_{ru}| \mu _{{\boldsymbol{\Delta }}^i_{ru}}^{\left( \xi  \right)},\delta _{{\boldsymbol{\Delta }}^i_{ru}}^{\left( \xi  \right)}} \right),
\end{aligned}
\end{equation}

${\boldsymbol{\mu }}_{{\boldsymbol{\varphi }}\left( {{\zeta _{ru}}} \right)}^{\left( \xi  \right)}$ is the mean vector of $\xi$-th distribution ${q^{\left( \xi  \right)}}\left( {{\boldsymbol{\varphi }}\left( {{\zeta _{ru}}} \right)} \right)$ and
\begin{equation}\label{variational_alpharu}
{q^{\left( \xi  \right)}}\left( {{\boldsymbol{\varphi }}\left( {{\zeta _{ru}}} \right)} \right) = \mathcal{CN}\left( {{\boldsymbol{\varphi }}\left( {{\zeta _{ru}}} \right)|{\boldsymbol{\mu }}_{{\boldsymbol{\varphi }}\left( {{\zeta _{ru}}} \right)}^{\left( \xi  \right)},{\boldsymbol{\Sigma }}_{{\boldsymbol{\varphi }}\left( {{\zeta _{ru}}} \right)}^{\left( \xi  \right)}} \right).
\end{equation}

Therefore, after simple manipulations, the expectation in \eqref{expectation_alphaau1} is given by
\begin{equation}\label{expectation_varphiru2_11}
\begin{aligned}
&{\mathbb{E}_{{q^{(\xi )}}\left( {{{\boldsymbol {\cal W}}_{\backslash {\alpha _{au}}}}} \right)}}\left( {\ln p\left( {{\boldsymbol{R}}|{\boldsymbol {\cal W}}} \right)} \right)\\
& =  - \frac{1}{2}\alpha _{au}^H\Gamma _{{\alpha _{au}}}^{\left( \xi  \right)}{\alpha _{au}} + \frac{1}{2}{\left( {\beta _{{\alpha _{au}}}^{\left( \xi  \right)}} \right)^H}{\alpha _{au}} + \frac{1}{2}\alpha _{au}^H\beta _{{\alpha _{au}}}^{\left( \xi  \right)} + {\cal C},
\end{aligned}
\end{equation}
where $\beta _{{\alpha _{au}}}^{\left( \xi  \right)} = \sum\limits_{t = 1}^T {\frac{1}{\delta }} \sqrt {{P_w}} {\left( {{\boldsymbol{\mu }}_{{\boldsymbol{\varphi }}\left( {{\zeta _{au}}} \right)}^{\left( \xi  \right)}} \right)^H}\left( {{\boldsymbol{r}}_t - {\boldsymbol{\Theta }}_{{\alpha _{ru}}}^{\left( {t,\xi } \right)}} \right)$ and  $\Gamma _{{ \alpha _{au}}}^{\left( \xi  \right)} = \frac{1}{\delta }T{P_w}\mathcal{B}_{{{\zeta}_{au}}}^{\left( {\xi } \right)}.$

By putting the prior distribution in \eqref{prior_distributionofPsis}, the another expectation term ${\mathbb{E}_{{{q^{(\xi )}}\left( {{{\boldsymbol {\cal W}}_{{\backslash {{\alpha _{au}}} }}}} \right)} }}\left( {\ln p\left( {{\alpha _{au}}} \right)} \right)$ is given by
\begin{equation}\label{expectation_prioralpha_au}
\begin{aligned}
 &{\mathbb{E}_{{{q^{(\xi )}}\left( {{{\boldsymbol {\cal W}}_{{\backslash {{\alpha _{au}}} }}}} \right)} }}\left( {\ln p\left( {{\alpha _{au}}} \right)} \right) \\&= -\frac{{\alpha _{au}^H{\alpha _{au}}}}{{{2\delta _{{{{\alpha }}_{au}}}}}} + \frac{{{\alpha ^H_{au}}{\mu _{{{{\alpha }}_{au}}}}}}{{{2\delta _{{\alpha_{au}}}}}} +  \frac{{{\alpha _{au}}{\mu ^H_{{{{\alpha }}_{au}}}}}}{{{2\delta _{{\alpha_{au}}}}}}+ {{\cal{C}}}.
\end{aligned}
\end{equation}

By inserting the equations \eqref{expectation_varphiru2_11} and \eqref{expectation_prioralpha_au} into \eqref{variaotionalalpha_au}, the $\xi$-th iteration variational distribution $q^{(\xi)}\left( {{{\alpha} _{au}}} \right)$ in \eqref{variaotionalalpha_au} can be given by

\begin{equation}\label{variaotionalalpha_au1}
{q^{\left( \xi  \right)}}\left( {{\alpha _{au}}} \right) \propto {\cal {CN}}\left( {{\alpha _{au}}}| {\mu^{\left(\xi\right)}_{{\alpha _{au}}}},{\delta^{\left(\xi\right)} _{{\alpha _{au}}}} \right),
\end{equation}
where  $\delta _{{\alpha _{au}}}^{\left( \xi  \right)} = {\left( {\Gamma _{{\boldsymbol \varphi _{\left( {{\zeta _{au}}} \right)}}}^{\left( \xi  \right)} + \frac{1}{{{\delta _{{\alpha _{au}}}}}}} \right)^{ - 1}}$ and $\mu _{{\alpha _{au}}}^{\left( \xi  \right)} = \delta _{{\alpha _{au}}}^{\left( \xi  \right)}\left( {\frac{{{\mu _{{\alpha _{au}}}}}}{{{\delta _{{\alpha _{au}}}}}} + \beta _{{\alpha _{au}}}^{\left( \xi  \right)}} \right)$.

\subsection{Estimation of Phase shift $\boldsymbol \varphi \left( {{\zeta _{au}}} \right)$}\label{audelay}
The line-of-sight time delay ${\zeta _{au}}$ is nonlinear to the phase shift term $\boldsymbol \varphi \left( {{\zeta _{au}}} \right)$ and it is difficult to directly estimate the time delay ${\zeta _{au}}$.  Hence, we first estimate the nonlinear LOS phase shift $\boldsymbol \varphi \left( {{\zeta _{au}}} \right)$.

 According to \eqref{variationalphik}, the $\xi$-th iteration variational distribution ${q^{(\xi )}}\left( {{\boldsymbol{\varphi }}\left( {{\zeta _{au}}} \right)} \right)$ can be formulated
\begin{equation}\label{variaotionalzeta_au}
\begin{aligned}
{q^{(\xi )}}\left( {{\boldsymbol{\varphi }}\left( {{\zeta _{au}}} \right)} \right) \propto \exp \left\{ {\mathbb{E}_{{q^{(\xi )}}\left( {{{\boldsymbol {\cal W}} _{\backslash {{\boldsymbol{\varphi }}\left( {{\zeta _{au}}} \right)}}}} \right)}}\left( {\ln p\left( {{\boldsymbol{R}}|{\boldsymbol {\cal W}}} \right)} \right) \right.
\\
\phantom{=\;\;}
\left.+  {\mathbb{E}_{{q^{(\xi )}}\left( {{{\boldsymbol {\cal W}} _{\backslash {{\boldsymbol{\varphi }}\left( {{\zeta _{au}}} \right)}}}} \right)}}\left( {\ln p\left( {{\boldsymbol{\varphi }}\left( {{\zeta _{au}}} \right)} \right)} \right) \right\}.
\end{aligned}
\end{equation}

Plugging the likelihood function \eqref{likelihoodnew}  into the first expectation term ${\mathbb{E}_{{q^{(\xi )}}\left( {{{\boldsymbol {\cal W}}_{\backslash {\varphi \left( {{\zeta _{au}}} \right)}}}} \right)}}\left( {\ln p\left( {{\boldsymbol{R}}|{\boldsymbol {\cal W}} } \right)} \right)$, it yields
\begin{equation}\label{expectation_zeta_au}
\begin{aligned}
&{\mathbb{E}_{{q^{(\xi )}}\left( {{{{\boldsymbol {\cal W}}}_{\backslash \varphi \left( {{\zeta _{au}}} \right)}}} \right)}}\left( {\ln p\left( {{\boldsymbol{R}}|{\boldsymbol {\cal W}}} \right)} \right) \\
 &=  - \frac{1}{{2\delta }}tr\left( {{\boldsymbol{\varphi }}\left( {{\zeta _{au}}} \right){{\boldsymbol{\varphi }}^H}\left( {{\zeta _{au}}} \right){\boldsymbol{\Gamma }}_{\varphi \left( {{\zeta _{au}}} \right)}^{\left( \xi  \right)} - 2{{\left( {{\boldsymbol{\beta }}_{\varphi \left( {{\zeta _{au}}} \right)}^{\left( \xi  \right)}} \right)}^H}{\boldsymbol{\varphi }}\left( {{\zeta _{au}}} \right)} \right) \\&+ \mathcal{C},
\end{aligned}
\end{equation}
where $\mathcal{C}$ is the terms irrelevant to the variable $ \varphi \left( {{\zeta _{au}}} \right)$ and can be regarded as constants. The other parameters are given by
\begin{equation}\label{app2}
{\boldsymbol{\beta }}_{{\boldsymbol{\varphi }}\left( {{\zeta _{au}}} \right)}^{\left( \xi  \right)} = \sum\limits_{t = 1}^T {\frac{1}{\delta }} {\left( {{\boldsymbol{r}}_t - {\boldsymbol{\Theta }}_{{\alpha _{ru}}}^{\left( {t,\xi } \right)}} \right)^H}\alpha^{\left( \xi  \right)}_{au}\sqrt {{P_w}} ,
\end{equation}
\begin{equation}\label{app1}
{\boldsymbol{\Gamma }}_{\varphi \left( {{\zeta _{au}}} \right)}^{\left( \xi  \right)} = T{P_w}\left(\left(\alpha^{\left( \xi  \right)}_{au}\right)^H\alpha^{\left( \xi  \right)}_{au}+ \delta _{{\alpha _{au}}}^{\left( \xi  \right)}\right),
\end{equation}
and ${\boldsymbol{\Theta }}_{{\alpha _{ru}}}^{\left( {t,\xi } \right)} = \sqrt {{P_w}} {\boldsymbol{\mu }}_{{\boldsymbol{\varphi }}\left( {{\zeta _{ru}}} \right)}^{\left( \xi  \right)}{{\boldsymbol{\Upsilon }}_t}\mathcal{A}{\boldsymbol{\mu }}_{{{\boldsymbol{\Delta }}_{ru}}}^{\left( \xi  \right)}$. ${\boldsymbol \mu}_{{{\boldsymbol \Delta}_{ru}}}^{\left( \xi  \right)}$ is  the mean of the variational distribution ${q^{\left( \xi  \right)}}\left( {{\boldsymbol{\Delta }}_{ru}} \right)$ and  is given by
\begin{equation}\label{variational_disDel}
\begin{aligned}
{q^{\left( \xi  \right)}}\left( {{\boldsymbol{\Delta }}_{ru}} \right) &\propto   {\cal C}{\cal N}\left( {{\boldsymbol{\Delta }}_{ru}|\boldsymbol \mu _{{\boldsymbol{\Delta }}_{ru}}^{\left( \xi  \right)},\boldsymbol \Sigma _{{\boldsymbol{\Delta }}_{ru}}^{\left( \xi  \right)}} \right)\\& = \prod\limits_{i=1}^{\text{PQ}}{\cal C}{\cal N}\left( {{\boldsymbol{\Delta }}^i_{ru}| \mu _{{\boldsymbol{\Delta }}^i_{ru}}^{\left( \xi  \right)},\delta _{{\boldsymbol{\Delta }}^i_{ru}}^{\left( \xi  \right)}} \right),
\end{aligned}
\end{equation}
and ${\boldsymbol{\mu }}_{{\boldsymbol{\varphi }}\left( {{\zeta _{ru}}} \right)}^{\left( \xi  \right)}$ is the mean vector of $\xi$-th distribution ${q^{\left( \xi  \right)}}\left( {{\boldsymbol{\varphi }}\left( {{\zeta _{ru}}} \right)} \right)$ and
\begin{equation}\label{variational_alpharu}
{q^{\left( \xi  \right)}}\left( {{\boldsymbol{\varphi }}\left( {{\zeta _{ru}}} \right)} \right) = \mathcal{CN}\left( {{\boldsymbol{\varphi }}\left( {{\zeta _{ru}}} \right)|{\boldsymbol{\mu }}_{{\boldsymbol{\varphi }}\left( {{\zeta _{ru}}} \right)}^{\left( \xi  \right)},{\boldsymbol{\Sigma }}_{{\boldsymbol{\varphi }}\left( {{\zeta _{ru}}} \right)}^{\left( \xi  \right)}} \right),
\end{equation}
will be given later.

Substituting \eqref{prior_distributionofPsit}  into the second expectation term ${\mathbb{E}_{{q^{(\xi )}}\left( {{\boldsymbol{\varphi }}\left( {{\zeta _{au}}} \right)} \right)}}\left( {\ln p\left( {{\boldsymbol{\varphi }}\left( {{\zeta _{au}}} \right)} \right)} \right)$, it yields
\begin{equation}\label{expectation_zetaau}
\begin{aligned}
&{\mathbb{E}_{{q^{(\xi )}}\left( {{\boldsymbol{\varphi }}\left( {{\zeta _{au}}} \right)} \right)}}\left( {\ln p\left( {{\boldsymbol{\varphi }}\left( {{\zeta _{au}}} \right)} \right)} \right)\\ &=  - \frac{1}{2}tr\left( {{\boldsymbol{\varphi }}^H}\left( {{\zeta _{au}}} \right){\boldsymbol{\Sigma }}_{{\boldsymbol{\varphi }}\left( {{\zeta _{au}}} \right)}^{ - 1}{\boldsymbol{\varphi }}\left( {{\zeta _{au}}} \right) \right.
\\
&\left.- 2{{\left( {{\boldsymbol{\mu }}_{{\boldsymbol{\varphi }}\left( {{\zeta _{au}}} \right)}^{\left( \xi  \right)}} \right)}^H}{\boldsymbol{\Sigma }}_{{\boldsymbol{\varphi }}\left( {{\zeta _{au}}} \right)}^{ - 1}{\boldsymbol{\varphi }}\left( {{\zeta _{au}}} \right) \right) + \mathcal{C},
\end{aligned}
\end{equation}

By substituting \eqref{expectation_zetaau} and \eqref{expectation_zeta_au} into \eqref{variaotionalzeta_au}, we can obtain
\begin{equation}\label{variaotionalzeta_au1}
{q^{\left( \xi  \right)}}\left( {{\boldsymbol{\varphi }}\left( {{\zeta _{au}}} \right)} \right) \propto {\cal C}{\cal N}\left( {{\boldsymbol{\varphi }}\left( {{\zeta _{au}}} \right)|{\boldsymbol{\mu }}_{{\boldsymbol{\varphi }}\left( {{\zeta _{au}}} \right)}^{\left( \xi  \right)},{\boldsymbol{\Sigma }}_{{\boldsymbol{\varphi }}\left( {{\zeta _{au}}} \right)}^{\left( \xi  \right)}} \right),
\end{equation}
where ${\boldsymbol{\mu }}_{{\boldsymbol{\varphi }}\left( {{\zeta _{au}}} \right)}^{\left( \xi  \right)}$ and ${\boldsymbol{\Sigma }}_{{\boldsymbol{\varphi }}\left( {{\zeta _{au}}} \right)}^{\left( \xi  \right)} $ are respectively given by
\begin{equation}\label{variaotionalzeta_au1_mean}
{\boldsymbol{\mu }}_{{\boldsymbol{\varphi }}\left( {{\zeta _{au}}} \right)}^{\left( \xi  \right)} = {\boldsymbol{\Sigma }}_{{\boldsymbol{\varphi }}\left( {{\zeta _{au}}} \right)}^{\left( \xi  \right)}\left( {{\boldsymbol{\mu }}_{{\boldsymbol{\varphi }}\left( {{\zeta _{au}}} \right)}^H{\boldsymbol{\Sigma }}_{{\boldsymbol{\varphi }}\left( {{\zeta _{au}}} \right)}^{ - 1} + {\boldsymbol{\beta }}_{{\boldsymbol{\varphi }}\left( {{\zeta _{au}}} \right)}^{\left( \xi  \right)}} \right),
\end{equation}
\begin{equation}\label{variaotionalzeta_au1_var}
{\boldsymbol{\Sigma }}_{{\boldsymbol{\varphi }}\left( {{\zeta _{au}}} \right)}^{\left( \xi  \right)} = {\left( {{\boldsymbol{\Gamma }}_{\varphi \left( {{\zeta _{au}}} \right)}^{\left( \xi  \right)} + {\boldsymbol{\Sigma }}_{{\boldsymbol{\varphi }}\left( {{\zeta _{au}}} \right)}^{ - 1}} \right)^{ - 1}}.
\end{equation}

\subsection{Estimation of Delay $\boldsymbol \varphi \left( {{\zeta _{ru}}} \right)$}\label{rudelay}
The estimation of reflecting path time delay ${\zeta _{ru}}$ is similar to the estimation of  $\boldsymbol \varphi \left( {{\zeta _{ru}}} \right)$ in subsection \ref{audelay}. According to \eqref{variationalphik}, the $\xi$-th iteration variational distribution ${q^{(\xi )}}\left(  \boldsymbol \varphi \left( {{\zeta _{ru}}} \right) \right)$ is given by
\begin{equation}\label{variaotionalzeta_ru}
\begin{aligned}
{q^{\left( \xi  \right)}}\left( \boldsymbol \varphi \left( {{\zeta _{ru}}} \right) \right) \propto \exp \left\{{\mathbb{E}_{{q^{(\xi )}}\left( {{{\boldsymbol {\cal W}}{\backslash {{\boldsymbol{\varphi }}\left( {{\zeta _{ru}}} \right)}}}} \right)}}\left( {\ln p\left( {{\boldsymbol{R}}|{\boldsymbol {\cal W}} } \right)} \right) \right.
\\
\phantom{=\;\;}
\left.+ {\mathbb{E}_{{q^{(\xi )}}\left( {{{\boldsymbol {\cal W}}_{\backslash {{\boldsymbol{\varphi }}\left( {{\zeta _{ru}}} \right)}}}} \right)}}\left( {\ln p\left( \boldsymbol \varphi \left( {{\zeta _{ru}}} \right) \right)} \right) \right\}.
\end{aligned}
\end{equation}

Substituting the likelihood function \eqref{likelihoodnew} into the first expectation term ${\mathbb{E}_{{q^{(\xi )}}\left( {{{\boldsymbol {\cal W}} _{\backslash {{\boldsymbol{\varphi }}\left( {{\zeta _{ru}}} \right)}}}} \right)}}\left( {\ln p\left( {{\boldsymbol{R}}|{\boldsymbol {\cal W}} } \right)} \right)$, it yields
\begin{equation}\label{expectation_zeta_au}
\begin{aligned}
&{\mathbb{E}_{{q^{(\xi )}}\left( {{{\boldsymbol {\cal W}} _{\backslash {\boldsymbol{\varphi }}\left( {{\zeta _{ru}}} \right)}}} \right)}}\left( {\ln p\left( {{\boldsymbol{R}}|{\boldsymbol {\cal W}}} \right)} \right)\\
& =  - \frac{1}{2}tr\left( {{\boldsymbol{\varphi }}\left( {{\zeta _{ru}}} \right){{\boldsymbol{\varphi }}^H}\left( {{\zeta _{ru}}} \right){\boldsymbol{\Gamma }}_{\boldsymbol \varphi \left( {{\zeta _{ru}}} \right)}^{\left( \xi  \right)} - 2{{\left( {{\boldsymbol{\beta }}_{{\boldsymbol{\varphi }}\left( {{\zeta _{ru}}} \right)}^{\left( \xi  \right)}} \right)}^H}{\boldsymbol{\varphi }}\left( {{\zeta _{ru}}} \right)} \right) \\ &+ \mathcal{C},
\end{aligned}
\end{equation}
where
with ${\boldsymbol \Gamma}_{\boldsymbol \varphi \left( {{\zeta _{ru}}} \right)}^{\left( \xi  \right)}$ and $\boldsymbol \beta _{\boldsymbol \varphi \left( {{\zeta _{ru}}} \right)}^{\left( \xi  \right)}$ are respectively given by
\begin{equation}\label{Tau_varphi}
\begin{aligned}
{\boldsymbol{\Gamma }}_{{\boldsymbol{\varphi }}\left( {{\zeta _{ru}}} \right)}^{\left( \xi  \right)} &= \sum\limits_{t = 1}^T {{\mathbb{E}_{{q^{(\xi )}}\left( {{{\boldsymbol {\cal W}}_{\backslash {\boldsymbol{\varphi }}\left( {{\zeta _{ru}}} \right)}}} \right)}}} \left( {\frac{{{P_w}}}{\delta }{{\boldsymbol{\Upsilon }}_t}{\cal{A}}{{\boldsymbol{\Delta }}_{ru}}{\boldsymbol{\Delta }}_{ru}^H{{\cal{A}}^H}{\boldsymbol{\Upsilon }}_t^H} \right) \\
&= \sum\limits_{t = 1}^T{\frac{{{P_w}}}{\delta }} {{\boldsymbol{\Upsilon }}_t}{\cal{A}}\left( {{\boldsymbol{\mu }}_{{{\boldsymbol{\Delta }}_{ru}}}^{\left( \xi  \right)}{{\left( {{\boldsymbol{\mu }}_{{{\boldsymbol{\Delta }}_{ru}}}^{\left( \xi  \right)}} \right)}^H} + {\boldsymbol{\Sigma }}_{{{\boldsymbol{\Delta }}_{ru}}}^{\left( \xi  \right)}} \right){{\cal{A}}^H}{\boldsymbol{\Upsilon }}_t^H,
\end{aligned}
\end{equation}
\begin{equation}\label{betavarphi}
\boldsymbol \beta _{\boldsymbol \varphi \left( {{\zeta _{ru}}} \right)}^{\left( \xi  \right)} = \sum\limits_{t = 1}^T  \frac{1}{\delta } \left( {\boldsymbol r_t - \boldsymbol \Theta _{{\alpha _{au}}}^{\left( \xi  \right)}} \right)\sqrt {P_w}\left({ {\boldsymbol{\mu }}_{{{\boldsymbol{\Delta }}_{ru}}}^{\left( \xi  \right)}}\right)^H  {\cal{A}}^H\boldsymbol \Upsilon^H_t .
\end{equation}
where ${\boldsymbol{\Theta }}_{{\alpha _{au}}}^{\left( \xi  \right)} = {\alpha ^{\left( \xi  \right)}_{au}}\sqrt {{P_w}} {\boldsymbol{\mu }}_{{\boldsymbol{\varphi }}\left( {{\zeta _{au}}} \right)}^{\left( \xi  \right)}$.

By putting the prior distribution \eqref{prior_distributionofPsit} into the second expectation term, the second expectation term can be given by
\begin{equation}\label{expectation_zeta_au2}
\begin{aligned}
&{\mathbb{E}_{{q^{(\xi )}}\left( {{\boldsymbol{\varphi }}\left( {{\zeta _{ru}}} \right)} \right)}}\left( {\ln p\left( {{\boldsymbol{\varphi }}\left( {{\zeta _{ru}}} \right)} \right)} \right) \\ &=  - \frac{1}{2}tr\left( {{{\boldsymbol{\varphi }}^H}\left( {{\zeta _{ru}}} \right){\boldsymbol{\Sigma }}_{{\boldsymbol{\varphi }}\left( {{\zeta _{ru}}} \right)}^{ - 1}{\boldsymbol{\varphi }}\left( {{\zeta _{ru}}} \right) - 2{\boldsymbol{\mu }}_{{\boldsymbol{\varphi }}\left( {{\zeta _{ru}}} \right)}^H{\bf{\Sigma }}_{{\boldsymbol{\varphi }}\left( {{\zeta _{ru}}} \right)}^{ - 1}{\boldsymbol{\varphi }}\left( {{\zeta _{ru}}} \right)} \right) \\ &+ \mathcal{C}.
\end{aligned}
\end{equation}

Plugging  \eqref{expectation_zeta_au} and \eqref{expectation_zeta_au2} into  \eqref{variaotionalzeta_ru} yields
\begin{equation}\label{expectation_zeta_au3}
{q^{\left( \xi  \right)}}\left( {{\boldsymbol{\varphi }}\left( {{\zeta _{au}}} \right)} \right) \propto {\cal C}{\cal N}\left( {{\boldsymbol{\varphi }}\left( {{\zeta _{au}}} \right)|{\boldsymbol{\mu }}_{{\boldsymbol{\varphi }}\left( {{\zeta _{au}}} \right)}^{\left( \xi  \right)},{\boldsymbol{\Sigma }}_{{\boldsymbol{\varphi }}\left( {{\zeta _{au}}} \right)}^{\left( \xi  \right)}} \right),
\end{equation}
where
\begin{equation}\label{expectation_zeta_rumu}
{\boldsymbol{\mu }}_{{\boldsymbol{\varphi }}\left( {{\zeta _{ru}}} \right)}^{\left( \xi  \right)} = {\boldsymbol{\Sigma }}_{{\boldsymbol{\varphi }}\left( {{\zeta _{ru}}} \right)}^{\left( \xi  \right)}\left( {{\boldsymbol{\mu }}_{{\boldsymbol{\varphi }}\left( {{\zeta _{ru}}} \right)}^H{\boldsymbol{\Sigma }}_{{\boldsymbol{\varphi }}\left( {{\zeta _{ru}}} \right)}^{ - 1} + {\boldsymbol{\beta }}_{{\boldsymbol{\varphi }}\left( {{\zeta _{ru}}} \right)}^{\left( \xi  \right)}} \right),
\end{equation}

\begin{equation}\label{expectation_zeta_ruvar}
{\boldsymbol{\Sigma }}_{{\boldsymbol{\varphi }}\left( {{\zeta _{ru}}} \right)}^{\left( \xi  \right)} = {\left( {{\boldsymbol{\Gamma }}_{\varphi \left( {{\zeta _{ru}}} \right)}^{\left( \xi  \right)} + {\boldsymbol{\Sigma }}_{{\boldsymbol{\varphi }}\left( {{\zeta _{ru}}} \right)}^{ - 1}} \right)^{ - 1}}.
\end{equation}

\subsection{Estimation of Inverse Variance ${w _{{\boldsymbol{\Delta }}_{ru}^i}}$}\label{wi}
According to \eqref{variationalphik}, the $\xi$-th iteration variational distribution $q^{(\xi)}\left( {w _{{\boldsymbol{\Delta }}_{ru}^i}} \right)$ is given by
\begin{equation}\label{variational_wi}
\begin{aligned}
{q^{\left( \xi  \right)}}\left( {{w_{{\boldsymbol{\Delta }}_{ru}^i}}} \right) \propto \exp \left\{{\mathbb{E}_{{q^{(\xi )}}\left( {{{\boldsymbol {\cal W}}_{\backslash {w_{{\boldsymbol{\Delta }}_{ru}^i}}}}} \right)}}\left({\ln p\left( {{\boldsymbol{\Delta }}_{ru}^i} \right)}\right)  \right.
\\
\phantom{=\;\;}
\left. +{\mathbb{E}_{{q^{(\xi )}}\left( {{{\boldsymbol {\cal W}}_{\backslash {w_{{\boldsymbol{\Delta }}_{ru}^i}}}}} \right)}}\left( {\ln p\left( {{w_{{\boldsymbol{\Delta }}_{ru}^i}}} \right)} \right)\right\}.
\end{aligned}
\end{equation}

Plugging \eqref{prior_Delta} into the first expectation term in  \eqref{variational_wi}, it yields
\begin{equation}\label{expectation_variational_omega}
\small
\begin{aligned}
&{\mathbb{E}_{{q^{(\xi )}}\left( {{{\boldsymbol {\cal W}}_{\backslash {w_{{\boldsymbol{\Delta }}_{ru}^i}}}}} \right)}}\left( {\ln p\left( {{\boldsymbol{\Delta }}_{ru}^i} \right)} \right) \\&= {\mathbb{E}_{{q^{(\xi )}}\left( {{{\boldsymbol {\cal W}}_{\backslash {w_{{\boldsymbol{\Delta }}_{ru}^i}}}}} \right)}}\left(\sum \limits_{l=1}^2 {{g_{i,l}}\ln \mathcal{CN}\left( {{\boldsymbol{\Delta }}_{ru}^i|{\mu^l_{{\boldsymbol{\Delta }}}},w_{{\boldsymbol{\Delta }}_{ru}^i}^{ - 1}} \right)} \right)\\
& = \frac{1}{2}\ln {w_{{\boldsymbol{\Delta }}_{ru}^i}} - \varpi^{\left( \xi  \right)}_i{w_{{\boldsymbol{\Delta }}_{ru}^i}},
\end{aligned}
\end{equation}
where $\varpi _i^{\left( \xi  \right)} = \frac{1}{2}\sum\limits_{l = 1}^2 {\hbar _{i,l}^{\left( \xi  \right)}} \left( {{{\left( {\mu _{{\boldsymbol{\Delta }}_{ru}^i}^{\left( \xi  \right)} - \mu _{{\boldsymbol{\Delta }}}^l} \right)}^2} + \delta _{{\boldsymbol{\Delta }}_{ru}^i}^{\left( \xi  \right)}} \right),$
and $\hbar _{i,l}^{\left( \xi  \right)}$ is the $\xi$-th estimated probability from the variational distribution ${q^{(\xi )}}\left( {{{\boldsymbol{g}}}} \right)$.

The prior distribution is assumed to follow a Gamma distribution in \eqref{Gamma_prior} and the expectation term ${\mathbb{E}_{{q^{(\xi )}}\left( {{{\boldsymbol {\cal W}}_{\backslash {w_{{\boldsymbol{\Delta }}_{ru}^i}}}}} \right)}}\left( {\ln p\left( {{w_{{\boldsymbol{\Delta }}_{ru}^i}}} \right)} \right)$ is given by
\begin{equation}\label{variational_expec_gamma}
{\mathbb{E}_{{q^{(\xi )}}\left( {{{\boldsymbol {\cal W}}_{\backslash {w_{{\boldsymbol{\Delta }}_{ru}^i}}}}} \right)}}\left( {\ln p\left( {{w_{{\boldsymbol{\Delta }}_{ru}^i}}} \right)} \right) = \left( {{a_i} - 1} \right)\ln {w_{{\boldsymbol{\Delta }}_{ru}^i}} - \frac{1}{{{b_i}}}{w_{{\boldsymbol{\Delta }}_{ru}^i}}.
\end{equation}

With the results \eqref{expectation_variational_omega}, \eqref{variational_expec_gamma}, the variational distribution $q^{(\xi)}\left( {w _{{\boldsymbol{\Delta }}_{ru}^i}} \right)$ can be given by
\begin{equation}\label{variational_gammas}
{q^{\left( \xi  \right)}}\left( {{w_{{\boldsymbol{\Delta }}_{ru}^i}}} \right) \propto \Gamma \left( {{w_{{\boldsymbol{\Delta }}_{ru}^i}}|a_i^{\left( \xi  \right)},b_i^{\left( \xi  \right)}} \right),
\end{equation}
where $a_i^{\left( \xi  \right)} = {a_i} + \frac{1}{2}$ and $b_i^{\left( \xi  \right)} = {\left( {\frac{1}{{{b_i}}} +  \frac{\varpi _i^{\left(\xi\right)}}{2}} \right)^{ - 1}}.$

Hence, the $\xi$-th estimation of ${w _{{\boldsymbol{\Delta }}_{ru}^i}}$ can be given by
\begin{equation}\label{estimation_wi}
{w^{\left(\xi\right)}_{{\boldsymbol{\Delta }}_{ru}^i}} = {{a_i^{\left( \xi  \right)}}}/{{b_i^{\left( \xi  \right)}}}.
\end{equation}
\subsection{Estimation of Sparse Vector ${{\boldsymbol{\Delta }}_{ru}}$}\label{deltai}
The sparse vector ${{\boldsymbol{\Delta }}_{ru}}$ is a one nonzero element vector and the nonzero element is the reflected path gain. Thus, the estimation of the sparse vector is equivalent to estimation of the  reflected path gain $\alpha_{ru}$. Meanwhile, the location of the nonzero element in the sparse vector ${{\boldsymbol{\Delta }}_{ru}}$ determines the true steering vector in \eqref{discretization}. Thus, the estimation of sparse vector ${{\boldsymbol{\Delta }}_{ru}}$ has key impacts on the localization and channel estimation performance.

According to \eqref{variationalphik}, the $\xi$-th iteration variational distribution $q^{(\xi)}\left( {{\boldsymbol{\Delta }}_{ru}} \right)$ is given by
\begin{equation}\label{variaotionalzeta_ru}
\begin{aligned}
{q^{\left( \xi  \right)}}\left( {{\boldsymbol{\Delta }}_{ru}} \right) \propto  \exp \left\{{\mathbb{E}_{{{q^{(\xi )}}\left( {{{\boldsymbol {\cal W}}_{ \backslash {{\boldsymbol{\Delta }}_{ru}}}}} \right)} }}\left( {\ln p\left( {{\boldsymbol{R}}|{\boldsymbol {\cal W}}} \right)} \right)  \right.
\\
\phantom{=\;\;}
\left. + {\mathbb{E}_{ {{q^{(\xi )}}\left( {{{\boldsymbol {\cal W}}_{ \backslash {{\boldsymbol{\Delta }}_{ru}}}}} \right)} }}\left( {\ln p\left({{\boldsymbol{\Delta }}_{ru}} \right)} \right)\right\}.
\end{aligned}
\end{equation}

Using similar steps, the first expectation  term can be given by
\begin{equation}\label{varaitional_Delta_ru1}
\begin{aligned}
&{\mathbb{E}_{{q^{(\xi )}}\left( {{{\boldsymbol {\cal W}}_{\backslash {{\boldsymbol{\Delta }}_{ru}}}}} \right)}}\left( {\ln p\left( {{\boldsymbol{R}}|{\boldsymbol {\cal W}}} \right)} \right)\\
&=  - \frac{1}{{2}}\left( {{\boldsymbol{\Delta }}_{ru}^H{\cal S}_{{{\boldsymbol{\Delta }}_{ru}}}^{\left( \xi  \right)}{{\boldsymbol{\Delta }}_{ru}} - {{\left( {\boldsymbol \beta _{{{\boldsymbol{\Delta }}_{ru}}}^{\left( \xi  \right)}} \right)}^H}{{\boldsymbol{\Delta }}_{ru}} - {\boldsymbol{\Delta }}_{ru}^H\boldsymbol \beta _{{{\boldsymbol{\Delta }}_{ru}}}^{\left( \xi  \right)}} \right) + {\cal C},
\end{aligned}
\end{equation}
where $\mathcal{S}_{{\boldsymbol{\Delta }}_{ru}}^{\left( \xi  \right)}$ and $\boldsymbol \beta _{{\boldsymbol{\Delta }}_{ru}}^{\left( \xi  \right)}$ are respectively given by
\begin{equation}\label{variational_s1}
\begin{aligned}
& \mathcal{S}_{{{\boldsymbol{\Delta }}_{ru}}}^{\left( \xi  \right)}\\ & = \sum\limits_{t = 1}^T {\frac{1}{\delta }} {\mathbb{E}_{{q^{(\xi )}}\left( {{{\boldsymbol {\cal W}}_{\backslash {{\boldsymbol{\Delta }}_{ru}}}}} \right)}}\left( {{P_w}{\mathcal{A}^H}{\boldsymbol{\Upsilon }}_t^H{{\boldsymbol{\varphi }}^H}\left( {{\zeta _{ru}}} \right){\boldsymbol{\varphi }}\left( {{\zeta _{ru}}} \right){{\boldsymbol{\Upsilon }}_t}\mathcal{A}} \right)\\
 & = \sum\limits_{t = 1}^T {\frac{1}{\delta }} \left( {{P_w}{\mathcal{A}^H}{\boldsymbol{\Upsilon }}_t^H{\cal B}_{{{\zeta}_{ru}}}^{\left( {\xi } \right)} {{\boldsymbol{\Upsilon }}_t}\mathcal{A}} \right)
\end{aligned},
\end{equation}
 where the parameter ${\boldsymbol{\beta }}_{{{\boldsymbol{\Delta }}_{ru}}}^{\left( \xi  \right)}$ is given by
\begin{equation}\label{variational_s2}
{\boldsymbol{\beta }}_{{{\boldsymbol{\Delta }}_{ru}}}^{\left( \xi  \right)} = \sum\limits_{t = 1}^T {\frac{{\sqrt {{P_w}} }}{\delta }} {{\cal A}^H}{\boldsymbol{\Upsilon }}_t^H{\left( {{\boldsymbol{\mu }}_{{\boldsymbol{\varphi }}\left( {{\zeta _{ru}}} \right)}^{\left( \xi  \right)}} \right)^H}\left( {{{\boldsymbol{r}}_t} - {\boldsymbol{\Theta }}_{{\alpha _{au}}}^{\left( \xi  \right)}} \right).
\end{equation}

Substituting the prior distribution in \eqref{prior_Delta} into the second expectation term, we can obtain
{\color{red}\begin{equation}\label{prior_refor}
\small
\begin{aligned}
&{\mathbb{E}_{{q^{(\xi )}}\left( {{{\boldsymbol{W}}_{\backslash {{\boldsymbol{\Delta }}_{ru}}}}} \right)}}\left( {\ln p\left( {{{\boldsymbol{\Delta }}_{ru}}} \right)} \right)\\
&= {\mathbb{E}_{{q^{(\xi )}}\left( {{{\boldsymbol{W}}_{\backslash {{\boldsymbol{\Delta }}_{ru}}}}} \right)}}\left( {\ln \left\{ {\prod\limits_{i = 1}^{{\rm{PQ}}} {\prod\limits_{l = 1}^2 {{\cal C}{\cal N}{{\left( {{\boldsymbol{\Delta }}_{ru}^i|\mu _{\boldsymbol{\Delta }}^l,w_{{\boldsymbol{\Delta }}_{ru}^i}^{ - 1}} \right)}^{{g_{i,l}}}}} } } \right\}} \right)\\
&=  \\ &- \frac{1}{2}\sum\limits_{i = 1}^{{\rm{PQ}}} {\sum\limits_{l = 1}^2 {{{\left( {{\boldsymbol{\Delta }}_{ru}^i - \mu _{\boldsymbol{\Delta }}^l} \right)}^H}{\mathbb{E}_{{q^{(\xi )}}\left( {{{\boldsymbol{W}}_{\backslash {{\boldsymbol{\Delta }}_{ru}}}}} \right)}}\left( {{g_{i,l}}{w_{{\boldsymbol{\Delta }}_{ru}^i}}} \right)\left( {{{\boldsymbol{\Delta }}_{ru}} - \mu _{\boldsymbol{\Delta }}^l} \right)} }\\  &+ {\cal C}\\
&=  - \frac{1}{2}\sum\limits_{l = 1}^2 {{{\left( {{{\boldsymbol{\Delta }}_{ru}} - {\boldsymbol{\mu }}_{\boldsymbol{\Delta }}^l} \right)}^H}{\mathbb{E}_{{q^{(\xi )}}\left( {{{\boldsymbol{W}}_{\backslash {{\boldsymbol{\Delta }}_{ru}}}}} \right)}}\left( {{{\boldsymbol{\Lambda }}_l}{\boldsymbol{\Sigma }}_{\bf{\Delta }}^{ - 1}} \right)\left( {{{\boldsymbol{\Delta }}_{ru}} - {\boldsymbol{\mu }}_{\boldsymbol{\Delta }}^l} \right) }\\ &+ {\cal C}
\end{aligned}
\end{equation}}
where $\boldsymbol{\Lambda}_l = diag\left\{ {{g_{1,l}}, \cdots ,{g_{{\text{PQ}},l}}} \right\}$, ${{\boldsymbol{\Sigma }}_{\boldsymbol{\Delta }}} = diag\left\{ { w_{\boldsymbol \Delta _{ru}^1} ^{ - 1}}, \cdots,{ w_{\boldsymbol \Delta _{ru}^{\text{PQ}}} ^{ - 1}}  \right\}$ and ${{\boldsymbol g}_l} = \left[ {{g_{1,l}}, \cdots ,{g_{\text{PQ},l}}} \right]$ and
\begin{equation}\label{variational_M}
\begin{aligned}
{\cal M}_{l}^{\left( \xi  \right)}& = {\mathbb{E}_{{q^{(\xi )}}\left( {{{\boldsymbol {\cal W}}_{\backslash {{\boldsymbol{\Delta }}_{ru}}}}} \right)}}\left( {\boldsymbol{\Lambda}_l{\boldsymbol{\Sigma }}_{\boldsymbol \Delta} ^{ - 1}} \right) \\ &= diag\left\{ {\frac{{a_1^{\left( \xi  \right)}}}{{b_1^{\left( \xi  \right)}}}\hbar _{1,l}^{\left( \xi  \right)}, \cdots ,\frac{{a_{\text{PQ}}^{\left( \xi  \right)}}}{{b_{\text{PQ}}^{\left( \xi  \right)}}}\hbar _{\text{PQ},l}^{\left( \xi  \right)}} \right\}.
\end{aligned}
\end{equation}

Substituting \eqref{variational_M} into \eqref{prior_refor}, the  expectation term ${\mathbb{E}_{{{q^{(\xi )}}\left( {{{\boldsymbol {\cal W}}_{\backslash {{\boldsymbol{\Delta }}_{ru}}}}} \right)} }}\left( {\ln p\left( {{\boldsymbol{\Delta }}_{ru}} \right) } \right)$ can be given by
\begin{equation}\label{variational_M1}
\begin{aligned}
&{\mathbb{E}_{{q^{(\xi )}}\left( {{{\boldsymbol {\cal W}}_{\backslash {{\boldsymbol{\Delta }}_{ru}}}}} \right)}}\left( {\ln p\left( {{{\boldsymbol{\Delta }}_{ru}}} \right)} \right) \\&=  - \frac{1}{2}\left( {{\boldsymbol{\Delta }}_{ru}^H{\boldsymbol{\Omega }}_{{{\boldsymbol{\Delta }}_{ru}}}^{\left( \xi  \right)}{{\boldsymbol{\Delta }}_{ru}} - {\boldsymbol{\Delta }}_{ru}^H{\boldsymbol{\varrho }}_{{{\boldsymbol{\Delta }}_{ru}}}^{\left( \xi  \right)} - {{\left( {{\boldsymbol{\varrho }}_{{{\boldsymbol{\Delta }}_{ru}}}^{\left( \xi  \right)}} \right)}^H}{{\boldsymbol{\Delta }}_{ru}}} \right) + \mathcal{C},
\end{aligned}
\end{equation}
where ${\boldsymbol{\Omega }}_{{{\boldsymbol{\Delta }}_{ru}}}^{\left( \xi  \right)} = \sum\limits_{l = 1}^2 {{\cal M}_l^{\left( \xi  \right)}}$ and ${\boldsymbol{\varrho }}_{{{\boldsymbol{\Delta }}_{ru}}}^{\left( \xi  \right)} = \sum\limits_{l = 1}^2 {{\cal M}_l^{\left( \xi  \right)}{\boldsymbol{\mu }}_{\boldsymbol{\Delta }}^l}.$

By plugging \eqref{varaitional_Delta_ru1} and \eqref{variational_M1} in \eqref{variaotionalzeta_ru},  the variational distribution $q^{(\xi)}\left( {{\boldsymbol{\Delta }}_{ru}} \right)$ is given by
\begin{equation}\label{variaotionalzeta_ruq}
\begin{aligned}
{q^{\left( \xi  \right)}}\left( {{\boldsymbol{\Delta }}_{ru}} \right) \propto {\cal C}{\cal N}\left( {{\boldsymbol{\Delta }}_{ru}|\boldsymbol \mu _{{\boldsymbol{\Delta }}_{ru}}^{\left( \xi  \right)},\boldsymbol \Sigma _{{\boldsymbol{\Delta }}_{ru}}^{\left( \xi  \right)}} \right),
\end{aligned}
\end{equation}
where $\boldsymbol \mu _{{\boldsymbol{\Delta }}_{ru}}^{\left( \xi  \right)}$ and $\boldsymbol \Sigma _{{\boldsymbol{\Delta }}_{ru}}^{\left( \xi  \right)}$ are respectively given by
\begin{equation}\label{Delta_var_q1}
{\boldsymbol{\mu }}_{{{\boldsymbol{\Delta }}_{ru}}}^{\left( \xi  \right)} = {\boldsymbol{\Sigma }}_{{{\boldsymbol{\Delta }}_{ru}}}^{\left( \xi  \right)}\left( {\boldsymbol \varrho _{{{\boldsymbol{\Delta }}_{ru}}}^{\left( \xi  \right)} + {\boldsymbol{\beta }}_{{{\boldsymbol{\Delta }}_{ru}}}^{\left( \xi  \right)}} \right),
\end{equation}
and
\begin{equation}\label{Delta_mean_q1}
{\boldsymbol{\Sigma }}_{{{\boldsymbol{\Delta }}_{ru}}}^{\left( \xi  \right)} = {\rm{ }}{\left( {\mathcal{S}_{{{\boldsymbol{\Delta }}_{ru}}}^{\left( \xi  \right)} + {\bf{\Omega }}_{{{\boldsymbol{\Delta }}_{ru}}}^{\left( \xi  \right)}} \right)^{ - 1}}.
\end{equation}

\subsection{Estimation of Indicator ${{\boldsymbol{g}}}$}\label{sec_ri}

The indicator variable ${{\boldsymbol{g}}}$ is directly involved with the sparse vector ${{\boldsymbol{\Delta }}_{ru}}$. The indicator variable indicates the location of the nonzero element in the sparse vector and the dependency is given in \eqref{indicator}.

According to \eqref{variationalphik}, the $\xi$-th iteration variational distribution $q^{(\xi)}\left( {\boldsymbol g} \right)$ is given by
\begin{equation}\label{variational_ri}
\begin{aligned}
{q^{\left( \xi  \right)}}\left( {{{\boldsymbol{g}}}} \right) \propto \exp \left\{{\mathbb{E}_{{q^{(\xi )}}\left( {{{\boldsymbol {\cal W}}_{\backslash {{\boldsymbol{g}}}}}} \right)}}\left( {\ln p\left( {{{\boldsymbol{g}}}|{\boldsymbol \chi }} \right)} \right) \right.
\\
\phantom{=\;\;}
\left. + {\mathbb{E}_{{q^{(\xi )}}\left( {{{\boldsymbol {\cal W}}_{\backslash {{\boldsymbol{g}}}}}} \right)}}\left( {\ln p\left( {{\boldsymbol{\Delta }}_{ru}} \right)} \right)\right\}.
\end{aligned}
\end{equation}

Substituting \eqref{multidis} into the first expectation term into \eqref{variational_ri}, we obtain the first expectation as
\begin{equation}\label{variational_r1}
{\mathbb{E}_{{q^{(\xi )}}\left( {{{\boldsymbol {\cal W}}_{\backslash {\boldsymbol{g}}}}} \right)}}\left( {\ln p\left( {{\boldsymbol{g}}|{\boldsymbol{\chi }}} \right)} \right){\rm{ = }}\sum\limits_{i = 1}^{\text{PQ}}\sum\limits_{l = 1}^{2} {{g_{i,l}}\ln {\chi _{i,l}}}.
\end{equation}

Following the similar steps in \eqref{prior_refor}, it yields
\begin{equation}\label{variational_r2}
\begin{aligned}
 &{\mathbb{E}_{{q^{(\xi )}}\left( {{{\boldsymbol {\cal W}}_{\backslash {\boldsymbol{g}}}}} \right)}}\left( {\ln p\left( {{{\boldsymbol{\Delta }}_{ru}}} \right)} \right)
 \\&= \sum\limits_{i = 1}^{\text{PQ}}\sum\limits_{l = 1}^{2} {{g_{i,l}}\ln \mathcal{CN}\left( {{\boldsymbol{\Delta }}_{ru}^i|{\mu^l _{{\boldsymbol{\Delta }}}},w_{{\boldsymbol{\Delta }}_{ru}^i}^{ - 1}} \right)}\\
 &= \sum\limits_{i = 1}^{\text{PQ}}\sum\limits_{l = 1}^{2}{g_{i,l}}  {\mathcal{Q}_{i,l}^{\left( \xi  \right)}}
\end{aligned},
\end{equation}
where ${\mu _{{\bf{\Delta }}_{ru}^i}^{\left( \xi  \right)}}$ is the $i$-th element in ${\boldsymbol{\mu }}_{{{\boldsymbol{\Delta }}_{ru}}}^{\left( \xi  \right)}$. $\mathcal{Q}_{i,l}^{\left( \xi  \right)} =  \frac{1}{2}\digamma \left( {a_i^{\left( \xi  \right)}} \right) + \frac{1}{2}\ln \left( {b_i^{\left( \xi  \right)}} \right) - \frac{1}{2}\left( {{{\left( {\mu _{{\boldsymbol{\Delta }}_{ru}^i}^{\left( \xi  \right)} - {\mu^l_{{\boldsymbol{\Delta }}}}} \right)}^2} + \delta _{{\boldsymbol{\Delta }}_{ru}^i}^{\left( \xi  \right)}} \right)a_i^{\left( \xi  \right)}/ b_i^{\left( \xi  \right)}$ and $\digamma \left(\cdot\right)$ is the digamma function.

Plugging \eqref{variational_r1} and \eqref{variational_r2} into \eqref{variational_ri} and considering the vector $\boldsymbol g$, it yields
\begin{equation}\label{variational_rdis}
{q^{(\xi )}}\left( {{{\boldsymbol{g}}}} \right) = \sum\limits_{i = 1}^{\text{PQ}}\sum\limits_{l = 1}^{2} \left({{\hbar _{i,l}^{\left( \xi  \right)}}}\right)^{{g_{i,l}}},
\end{equation}
where $\hbar _{i,l}^{\left( \xi  \right)} = \frac{\exp\left({ \mathcal{\bar Q}_{i,l}^{\left( \xi  \right)}}\right)}{{\sum\limits_{l = 1}^{2} {\exp\left({ \mathcal{\bar Q}_{i,l}^{\left( \xi  \right)}}\right)} }}$ and $\mathcal{\bar Q}_{i,l}^{\left( \xi  \right)} = \mathcal{Q}_{i,l}^{\left( \xi  \right)} + \ln {\chi _{l}}.$

\subsection{Estimation of UE Location}\label{user}
From the likelihood function in \eqref{likelihood}, it is computationally prohibitive to directly estimate the location of the user.  Fortunately, the estimated location of the user can be given by ${{\boldsymbol p}^{\left( \xi  \right)}_u} = {{\boldsymbol p}_r} + \rho{\boldsymbol \eta}^{\left( \xi  \right)}$, where $\rho$ is the distance between the UE and ${{\boldsymbol p}_r}$ to be estimated.
The vector ${\boldsymbol \eta}^{\left( \xi  \right)} =  - \frac{{2\pi }}{\lambda }{\left[ {\sin {\phi}^{\left( \xi  \right)} \cos {\psi}^{\left( \xi  \right)},\sin {\phi}^{\left( \xi  \right)} \sin {\psi}^{\left( \xi  \right)} ,\cos {\phi}^{\left( \xi  \right)} } \right]^T}$ is the waveform vector with the estimated azimuth and elevation angles.  Moreover, the location follows the geometric constraints, which can be given by
\begin{equation}\label{geocons1}
\begin{aligned}
\left( {{{\zeta }^{\left( \xi  \right)}_{ru}} - {{\zeta }^{\left( \xi  \right)}_{au}}} \right)c & = \underbrace {{{\left\| {\rho{\boldsymbol \eta}^{\left( \xi  \right)} } \right\|}_2}}_{{d^{\left( \xi  \right)}_{ru}}} + \underbrace {{{\left\| {{{\boldsymbol p}_a} - {{\boldsymbol p}_r}} \right\|}_2}}_{{d^{\left( \xi  \right)}_{ar}}} - \\ & \underbrace {{{\left\| {{{\boldsymbol p}_a} - {{\boldsymbol p}_r} - \rho {\boldsymbol \eta}^{\left( \xi  \right)}} \right\|}_2}}_{{d^{\left( \xi  \right)}_{au}}},
\end{aligned}
\end{equation}
where ${{\zeta }^{\left( \xi  \right)}_{ru}}$ and ${{\zeta }^{\left( \xi  \right)}_{au}}$ are the $\xi$-th estimation delays respectively. The delays ${\zeta _{au}}$ and ${{\zeta _{ru}}}$ can be estimated from $\boldsymbol \mu _{\boldsymbol \varphi \left( {{\zeta _{au}}} \right)}^{\left( \xi  \right)}$ and $\boldsymbol \mu _{\boldsymbol \varphi \left( {{\zeta _{ru}}} \right)}^{\left( \xi  \right)}$ respectively by following the results in \cite{JointWen18TVT}
\begin{equation}\label{timedelaymini1}
{{\hat \zeta }^{\left( \xi  \right)}_{ru}} = \arg \mathop {\min }\limits_{{\zeta _{ru}}} {\left\| {{{\left( {{\boldsymbol{\mu }}_{{\boldsymbol{\varphi }}\left( {{\zeta _{ru}}} \right)}^{\left( \xi  \right)}} \right)}^H}{\boldsymbol{\mu }}_{{\boldsymbol{\varphi }}\left( {{\zeta _{ru}}} \right)}^{\left( \xi  \right)}} \right\|_2},
\end{equation}
and
\begin{equation}\label{timedelaymini2}
{{\hat \zeta }^{\left( \xi  \right)}_{au}} = \arg \mathop {\min }\limits_{{\zeta _{au}}} {\left\| {{{\left( {{\boldsymbol{\mu }}_{{\boldsymbol{\varphi }}\left( {{\zeta _{au}}} \right)}^{\left( \xi  \right)}} \right)}^H}{\boldsymbol{\mu }}_{{\boldsymbol{\varphi }}\left( {{\zeta _{au}}} \right)}^{\left( \xi  \right)}} \right\|_2}.
\end{equation}

Hence, the estimation of $\rho$ can be obtained by minimizing
\begin{equation}\label{leastsquare}
\begin{aligned}
{\rho ^{\left( \xi  \right)}} = \arg \mathop {\min }\limits_\rho  {\left\| {\left( {\zeta _{ru}^{\left( \xi  \right)} - \zeta _{au}^{\left( \xi  \right)}} \right)c - d_{ru}^{\left( \xi  \right)} - d_{ar}^{\left( \xi  \right)} + d_{au}^{\left( \xi  \right)}} \right\|_2}.
 \end{aligned}
\end{equation}

By taking derivative with respect to $\rho$ and tedious manipulations, the solution $ \rho^{\left( \xi  \right)}$ is given by
\begin{equation}\label{solution_rho}
\small
\begin{aligned}
{\rho ^{\left( \xi  \right)}} = \frac{{{{\left( {\left( {\zeta _{ru}^{\left( \xi  \right)} - \zeta _{au}^{\left( \xi  \right)}} \right)c - d_{ar}^{\left( \xi  \right)}} \right)}^2} - {{\left( {d_{ar}^{\left( \xi  \right)}} \right)}^2}}}{{2\left( {\left( {\zeta _{ru}^{\left( \xi  \right)} - \zeta _{au}^{\left( \xi  \right)}} \right)c - d_{ar}^{\left( \xi  \right)}} \right)\left\| {{{\left( {{{\boldsymbol{\eta }}^{\left( \xi  \right)}}} \right)}^T}{{\boldsymbol{\eta }}^{\left( \xi  \right)}}} \right\| - 2{{\left( {{{\boldsymbol{p}}_a} - {{\boldsymbol{p}}_r}} \right)}^T}{{\boldsymbol{\eta }}^{\left( \xi  \right)}}}}.
\end{aligned}
\end{equation}

Hence, the $\xi$-th estimation of the UE location is given by
\begin{equation}\label{UEestimation}
{{\boldsymbol p}^{\left( \xi  \right)}_u} = {{\boldsymbol p}_r} + \rho^{\left( \xi  \right)} {\boldsymbol \eta}^{\left( \xi  \right)}.
\end{equation}

The user location and channel estimation involves various parameter estimation. For better and clearer presentation,  we summarize the JCLE algorithm as {\textbf {Algorithm I}} and implementation interpretations are given by
\begin{itemize}
  \item The time of flight (ToF) between the AP and RIS is hidden in the phase shift $\boldsymbol \varphi\left(\zeta_{au}\right)$. The ToF is estimated in subsections \ref{audelay};
  \item Similarly, the ToF between the RIS and the user is estimated in subsections \ref{rudelay};
  \item The indicator  $\boldsymbol g$ is a key parameter that directly determines the angles of arrival and it is estimated in \ref{sec_ri};
  \item Given the estimated angles and ToFs, the user location can be determined in \ref{user}.
  \item Nuisance parameters estimation are necessary to be included in the algorithm.
\end{itemize}

Our proposed algorithm is an iterative algorithm developed to approximate the true posterior distribution via mean-field factorization, KL divergence minimization, and alternating optimization. The convergence of the proposed variational Bayesian inference algorithm has been proven to converge \cite{TCRobustLi2020, fox2012tutorial}.

\section{Discussions}
{\color{red}In the paper,  channel estimation and localization share commonalities in their reliance on received signal parameters and the utilization of signal processing techniques. Both processes involve extracting meaningful information from the transmitted signals to achieve their respective goals.  For instance, the channel gains $\alpha_{au}$ and $\alpha_{ru}$, delays $\zeta_{au}$ and $\zeta_{ru}$, and angles $\phi$ and $\psi$ are often used in both channel estimation and localization algorithms. However, they have distinct objectives: channel estimation focusing on characterizing the communication channel, and localization aiming to determine spatial locations. Their interdependence on shared signal characteristics highlights the synergy between these two vital components in wireless communication systems.}

In {\textbf {Algorithm I}}, the complexity of the proposed algorithm mainly comes from the inversion in the covariance matrix when estimating of the sparse vector ${{{\boldsymbol{\Delta }}_{ru}}}$ in \eqref{Delta_mean_q1} in each iteration and other parameter estimation only involves scalars. The covariance matrix ${\boldsymbol{\Sigma }}_{{{\boldsymbol{\Delta }}_{ru}}}^{\left( \xi  \right)}$ is with the dimension of $PQ \times PQ$ and the inversion will involve computational complexity of $\mathcal{O}\left(\left(PQ\right)^3\right)$.  First, the matrix $\mathcal{S}_{{{\boldsymbol{\Delta }}_{ru}}}^{\left( \xi  \right)}$ can be reformulated as
\begin{equation}\label{newr}
\begin{aligned}
\mathcal{S}_{{{\boldsymbol{\Delta }}_{ru}}}^{\left( \xi  \right)} &= \sum\limits_{t = 1}^T {\frac{1}{\delta }} \left( {{P_w}{\mathcal{A}^H}{\boldsymbol{\Upsilon }}_t^H{\cal B}_{\zeta_{ru}}^{\left( {\xi } \right)} {{\boldsymbol{\Upsilon }}_t}\mathcal{A}} \right)
&= {\boldsymbol F}^H{\boldsymbol F},
\end{aligned}
\end{equation}
with ${\boldsymbol F} = \sqrt{\frac{P_w{{\cal B}_{{\zeta_{ru}}}^{\left( {\xi } \right)}}}{ {\delta}}} {{\boldsymbol{\Upsilon }}}\mathcal{A}$ and ${{\boldsymbol{\Upsilon }}}  = \left[{{\boldsymbol{\Upsilon }_1}}, {{\boldsymbol{\Upsilon }_2}},...,{{\boldsymbol{\Upsilon }_T}} \right]$.

Substituting \eqref{newr} into \eqref{Delta_mean_q1} and utilizing the matrix inverse lemma, we can obtain
\begin{equation}\label{woodcova}
\small
\begin{aligned}
{\boldsymbol{\Sigma }}_{{{\boldsymbol{\Delta }}_{ru}}}^{\left( \xi  \right)} &=
{\left( {{\boldsymbol F}^H{\boldsymbol F} + {{\boldsymbol \Omega }}_{{{\boldsymbol{\Delta }}_{ru}}}^{\left( \xi  \right)}} \right)^{ - 1}}\\
 &= \left({\boldsymbol \Omega}_{{{\boldsymbol{\Delta }}_{ru}}}^{\left( \xi  \right)}\right)^{-1} \\ &- \left(\boldsymbol{\Omega}_{{{\boldsymbol{\Delta }}_{ru}}}^{\left( \xi  \right)}\right)^{-1}{\boldsymbol F}^H\left( {\bf I}^{-1} + \left({\boldsymbol \Omega}_{{{\boldsymbol{\Delta }}_{ru}}}^{\left( \xi  \right)}\right)^{-1}\right)^{-1}{\boldsymbol F}\left({\boldsymbol \Omega}_{{{\boldsymbol{\Delta }}_{ru}}}^{\left( \xi  \right)}\right)^{-1},
\end{aligned}
\end{equation}
The computational complexity of estimating covariance matrix ${\boldsymbol{\Sigma }}_{{{\boldsymbol{\Delta }}_{ru}}}^{\left( \xi  \right)}$ in \eqref{woodcova} is reduced to $\mathcal{O}\left(TP^2Q^2\right)$.
 Hence, the total computational complexity of {\textbf{Algorithm I}} is proportional to  $\mathcal{O}\left(TP^2Q^2\right)$. The complexity of the PSO  algorithm is given by ${{\mathcal{O}}}\left( {2{{LTS}}\eta } \right)$, where $S$ and $\eta$ are the particle number and the convergence iterations. The complexity of the ML algorithm mainly comes from the IFFT-based time delay estimation, which has the complexity of ${\cal{O}}\left( {{{LT}}\log \left( {LT} \right) + MN\log \left( {MN} \right)} \right)$. Although the complexity of the proposed algorithm is possibly higher than that of the PSO and ML algorithms,  our algorithm can achieve better localization performance, which will be demonstrated in the simulation section.

{\begin{algorithm}[htbp]
  \caption{Variational Bayesian Inference-Based Localization and Channel Estimation Algorithm}
  \begin{algorithmic}[1]
      \State Input the distributions $p\left(\alpha_{au}\right)$, $p\left(\boldsymbol \varphi\left(\zeta_{au}\right)\right)$, $p\left(\boldsymbol \varphi\left(\zeta_{ru}\right)\right)$, $p\left( {\boldsymbol \Delta}_{ru} \right)$;
      \State Input the parameters $\delta$, ${P_w}$, $T$, $N$, $M$, $P$, $Q$, ${\boldsymbol p}_a$, ${\boldsymbol p}_u$, ${\boldsymbol p}_r$ and randomly generated $\boldsymbol \omega$, ${\mathcal T}$ and collect measurements $\boldsymbol R$;
      \State $\xi = 1$;
      \While {$|{\text{KL}}\left( {q^{\left(\xi+1\right)}\left( {\boldsymbol {\cal W}} \right)||p\left( {\boldsymbol {\cal W}} | {\boldsymbol R} \right)} \right) - {\text{KL}}\left( {q^{\left(\xi\right)}\left(  {\boldsymbol {\cal W}} \right)||p\left(  {\boldsymbol {\cal W}} | {\boldsymbol R}\right)} \right)|> {\mathcal T}$}
      \State Updating the mean ${\mu^{\left(\xi\right)}_{{\alpha _{au}}}}$ and variance $\delta _{{\alpha _{au}}}^{\left( \xi  \right)}$ of ${q^{\left( \xi  \right)}}\left( {{\alpha _{au}}} \right)$ via \eqref{variaotionalalpha_au1} respectively;
      \State Updating the mean ${\boldsymbol{\mu }}_{{\boldsymbol{\varphi }}\left( {{\zeta _{au}}} \right)}^{\left( \xi  \right)}$ and variance $ {\boldsymbol{\Sigma }}_{{\boldsymbol{\varphi }}\left( {{\zeta _{au}}} \right)}^{\left( \xi  \right)}$ of ${q^{\left( \xi  \right)}}\left({{\boldsymbol{\varphi }}\left( {{\zeta _{au}}} \right)}\right)$ via \eqref{variaotionalzeta_au1_mean} and \eqref{variaotionalzeta_au1_var} respectively;
      \State Updating the mean ${\boldsymbol{\mu }}_{{\boldsymbol{\varphi }}\left( {{\zeta _{ru}}} \right)}^{\left( \xi  \right)}$ and variance $ {\boldsymbol{\Sigma }}_{{\boldsymbol{\varphi }}\left( {{\zeta _{ru}}} \right)}^{\left( \xi  \right)}$ of ${q^{\left( \xi  \right)}}\left({{\boldsymbol{\varphi }}\left( {{\zeta _{ru}}} \right)}\right)$ via \eqref{expectation_zeta_rumu} and \eqref{expectation_zeta_ruvar} respectively;
      \State Updating the mean $ {\boldsymbol{\mu }}_{{{\boldsymbol{\Delta }}_{ru}}}^{\left( \xi  \right)}$ and variance $ {\boldsymbol{\Sigma }}_{{{\boldsymbol{\Delta }}_{ru}}}^{\left( \xi  \right)}$ of ${q^{\left( \xi  \right)}}\left( {{\boldsymbol{\Delta }}_{ru}} \right)$ via \eqref{Delta_var_q1} and \eqref{Delta_mean_q1} respectively;
      \State Updating the parameters $\hbar _{i,l}^{\left( \xi  \right)}$ and $\mathcal{\bar Q}_{i,l}^{\left( \xi  \right)}$ of ${q^{(\xi )}}\left( {{{\boldsymbol{r}}}} \right)$ via \eqref{variational_rdis} respectively;
      \State Updating the location ${{\boldsymbol p}^{\left( \xi  \right)}_u}$ via  $\eqref{UEestimation}$;
      \State Output the $\xi$-th estimation of  $\alpha_{au}$, ${{{\zeta _{au}}}}$, ${{{\zeta _{ru}}}}$, ${\boldsymbol \Delta}^i_{ru}$, ${\boldsymbol g}_i$, $\boldsymbol \chi$ and UE location ${\boldsymbol p}_u$ via \eqref{variaotionalzeta_au1_mean}, \eqref{expectation_zeta_rumu}, \eqref{Delta_var_q1}, $\chi _l^{\left( \xi  \right)} = \frac{{\lambda _l^{\left( \xi  \right)}}}{{\sum\limits_{l = 1}^2 {\lambda _l^{\left( \xi  \right)}} }}$,  \eqref{variational_rdis},  \eqref{UEestimation}  respectively;
      \State $\xi = \xi + 1;$
      \EndWhile
    \label{code:recentEnd}
  \end{algorithmic}
\end{algorithm}
}

\section{Simulation Results}
\subsection{Numerical Settings}
In this section, we investigate the estimation performance of the proposed algorithm in different scenarios. Given the multiple channel parameters, the sparse vector $\boldsymbol\Delta_{ru}$ estimation error and the angles estimation errors are presented to show the channel learning performance.
We consider a 3D localization and channel estimation of a user with the aid of the RIS system. {\color{red}The parameter settings are summarized as:
\begin{table}[H]
\caption{{\color{red}Parameter settings in simulation results}}
\centering
\begin{tabular}{|c|c|}
\hline
Parameter                               & Value                                                                  \\ \hline
${\boldsymbol p}_a$ &$ \left[100,100,30\right]^T$ \\ \hline
${\boldsymbol p}_r$ & $\left[10,40,10\right]^T$   \\ \hline
$M$                                       & 20                                                                           \\ \hline
$N$                                       & 20                                                                           \\ \hline
$P_w$                                    & 1W                                                                           \\ \hline
$T$                                       & 80                                                                           \\ \hline
$L$                                       & 128                                                                          \\ \hline
$P$                                       & 10                                                                           \\ \hline
$Q$                                       & 10                                                                           \\ \hline
$\delta$                                 & 0.01                                                                   \\ \hline
\end{tabular}
\end{table}}
The distance between the RIS elements is half wavelength. The angle of arrivals $\psi$ and $\phi$ range in $\left[-\frac{\pi}{2},\frac{\pi}{2}\right]$.  The  prior parameters of unnormalized channel gain $\alpha_{au}$ is given by $\mu_{\alpha_{au}} = 0.2+ 0.2i$ and $\delta_{\alpha_{au}} = 0.01$. For the sparse vector ${\boldsymbol{\Delta }}$, the means are given by ${\mu^1_{{\boldsymbol{\Delta }}}} = 0$, ${\mu^2_{{\boldsymbol{\Delta }}}} = 0.5+0.5i$. ${{\mathbb{E}}}\left( {w_{{\boldsymbol{\Delta }}_{ru}^i}^{ - 1}} \right){{ = }}{{{a_i}}}{{{b_i}}}$, $a_i = 10^5$ and $b_i = 10^{-3}$. The initial positions of the user for all presented algorithms are both generated by adding Gaussian distributed bias ${\mathcal{N}_p}\left( {{\boldsymbol{0}},{{{\boldsymbol{\tilde \Sigma }}}_p}} \right)$ to the true position ${\boldsymbol p}_{r}$ and ${{{\boldsymbol{\tilde \Sigma }}}_p} = 25\mathbf{I}_3$.  This prior information can be obtained via coarse estimation.The mentioned settings are unaltered and otherwise stated differently. For better clarification, the proposed algorithm is compared to the following algorithms:
\begin{itemize}
  \item The quasi-Newton and maximum likelihood estimator were proposed for the localization problem in a RIS-aided localization system with the perfect channel information \cite{KeykhosraviSISO21ICC};
  \item A PSO algorithm was proposed to tackle the optimization problem for the RIS-aided localization in \cite{MetaLocalizationZhangTWC21}.  As a search algorithm, PSO can find the local optimum solutions of the proposed problem.
  \item Bayesian Cramer-Rao lower bound (BCRB): BCRB is adopted here as a benchmark for evaluating the performance of the proposed algorithm and the benchmark is derived using the Fisher information matrix. The original unknown parameter vector $\boldsymbol  \phi$ involves the other nuisance parameters and the Fisher information matrix  is derived as \cite{TCRobustLi2020,Li21WCLRobust}
      \begin{equation}\label{newFIM}
      \small
\begin{aligned}
J\left( {\boldsymbol{\cal W}} \right)
& =   \sum\limits_{t = 1}^T {{\mathbb{E}_{p\left( {{\boldsymbol{R,}}{\boldsymbol{\cal W}}} \right)}}\left[ {{{\left( {\frac{{\partial \ln p\left( {\boldsymbol R|{\boldsymbol{\cal W}}} \right)}}{{\partial {\boldsymbol{\cal W}}}}} \right)}^H}\frac{{\partial \ln p\left( {\boldsymbol R|{\boldsymbol{\cal W}}} \right)}}{{\partial {\boldsymbol{\cal W}}}}} \right]} \\ &+   \sum\limits_{t = 1}^T {{{\mathbb{E}_{p\left( {{\boldsymbol{R,}}{\boldsymbol{\cal W}}} \right)}}\left[ {{{\left( {\frac{{\partial \ln p\left( {\boldsymbol{\cal W}} \right)}}{{\partial {\boldsymbol{\cal W}}}}} \right)}^H}\frac{{\partial \ln p\left( {\boldsymbol{\cal W}} \right)}}{{\partial {\boldsymbol{\cal W}}}}} \right]}} \\
& \approx \frac{1}{\sigma }\sum\limits_{t = 1}^T {{{\mathbb{E}_{p\left( {{\boldsymbol{R,}}{\boldsymbol{\cal W}}} \right)}}\left[ {\Re \left( \frac{{\partial {\boldsymbol{\Xi }}_t^H}}{{\partial {{\boldsymbol{\cal W}}^H}}}\frac{{\partial {{\boldsymbol{\Xi }}_t}}}{{\partial {\boldsymbol{\cal W}}}} \right)} \right]}}\\
\end{aligned},
\end{equation}
where ${{\boldsymbol{\Xi }}_t} = {{\boldsymbol{\Xi }}_{au}} + {\boldsymbol{\Xi }}_{ru}^t$. The Fisher information matrix in $\eqref{newFIM}$ implies that the LOS and RIS signals both contribute to the estimation of channel gains and the angles. The prior distributions does not directly involve the UE position and can be ignored.
For the localization error bound, we transform the unknown vector $ {\boldsymbol {\cal W}}$ to an equivalent unknown variable vector  ${\boldsymbol{\kappa }} = \left[ {{{\boldsymbol{p}}^T_u},{\zeta _{ru}},{\zeta_{au}},\psi ,\phi } \right]$ and the corresponding Fisher information matrix is given by \cite{TITShen10Fundamental}
      \begin{equation}\label{newFIM1}
     J\left( {\boldsymbol{\kappa }} \right) =  {\boldsymbol{\mathcal P}}J\left(  \boldsymbol  {\boldsymbol {\cal W}} \right){{\boldsymbol{\mathcal P}}^T},
      \end{equation}
      and ${\boldsymbol{\mathcal P}}$ is the transform matrix and given by ${\boldsymbol{\mathcal P}}= \frac{{\partial {{ \boldsymbol {\cal W}}}}}{{\partial {\boldsymbol \kappa ^T}}}$.    Hence the equivalent Cramer-Rao lower bound of UE position is given by
      \begin{equation}\label{CRLB}
      \text{BCRB}{_{{{{\boldsymbol p}}_u}}} \ge tr\left( {{{\left[ {{J^{ - 1}}\left( {\boldsymbol {\kappa }} \right)} \right]}_{1:3,1:3}}} \right).
      \end{equation}
      where ${\left[ \bullet \right]_{1:3,1:3}}$ is a block matrix composing of the first $3 \times 3$ elements in  $J\left( {\boldsymbol{\kappa }} \right)$. The detailed derivations of $J\left( \boldsymbol \kappa \right)$ are given in Appendix A.

      Using the results in \eqref{Jkappa}, \eqref{F} and  \eqref{D} in Appendix A, we can obtain
      \begin{equation}\label{bound}
       {\text{BCRB}}{_{{{\boldsymbol{p}}_u}}}\ge \sqrt{ tr\left[ {{{\left( {{{\boldsymbol{J}}_{11}} - {{\boldsymbol{J}}_{12}}{\boldsymbol{J}}_{22}^{ - 1}{\boldsymbol{J}}_{12}^T} \right)}^{ - 1}}} \right]}.
      \end{equation}
\end{itemize}

\subsection{Far Field Scenarios}\label{guo}
 In this subsection, the numerical results of our algorithm in the far-field channel estimation and localization problem are investigated. The true user location is set to meet the constraint ${\left\| {{{\boldsymbol p}_u} - {{\boldsymbol p}_r}} \right\|_2} = 20 >100\lambda$ and $\lambda$ is carrier wavelength.

\begin{figure}
\vspace{-2em}
  \centering
  \includegraphics[width=3.0in]{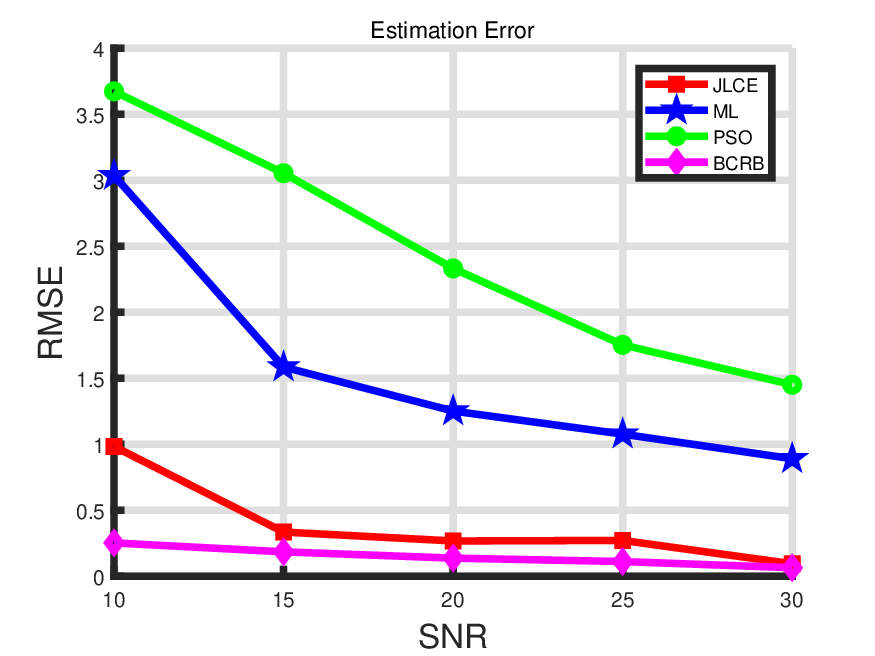}\\
  \caption{Localization performances with OFDM subcarrier number $L=128$ and snapshot $T=80$ under different SNRs  }\label{fig:FigforRISSNR}

\end{figure}

\begin{figure}
  \centering
  \includegraphics[width=3.0in]{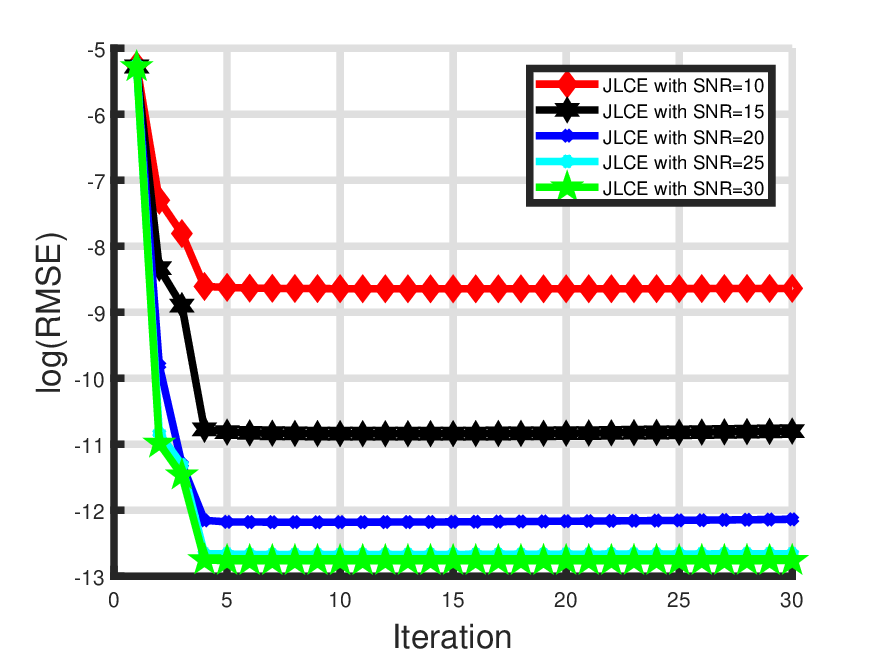}\\
  \caption{Sparse channel vector $\boldsymbol \Delta_{ru}$ estimation results with $L=128$ and snapshot $T=80$ under different SNRs }\label{fig:FigforRISSNR2}
    \vspace{-1em}
\end{figure}

In Fig.1, we first investigate the impact of signal-noise ratio (SNR) on the localization performance and estimation accuracy of the sparse vector $\boldsymbol \Delta_{ru}$. The PSO and ML algorithms both require the perfect knowledge of the reflected path channel gain $\alpha_{ru}$ and only the mean value of $\alpha_{ru}$ is available in this scenario.  The particle number of the PSO algorithm is set to be $200$. It is clear that the proposed algorithm JLCE can approach the BCRB with high SNR and the localization performance of the proposed algorithm JLCE outperforms the other algorithms. Because the proposed algorithm adopts the joint minimum mean square error (MMSE) estimation scheme and achieves accurate estimation of the sparse vector $\boldsymbol \Delta_{ru}$ and $\boldsymbol g$.  Meanwhile, the estimation performance of the sparse vector $\boldsymbol \Delta_{ru}$ under different SNRs is also investigated in Fig.2.  The proposed algorithm can achieve stable convergence at a rapid rate (less than $5$ iterations). The results in Fig.\ref{fig:FigforRISSNR} and Fig.\ref{fig:FigforRISSNR2} both intuitively show that the localization and estimation performances will increase with the higher SNR and the proposed algorithm can achieve better performances.


\begin{figure}
  \vspace{-0.2em}
  \centering
  \includegraphics[width=3.0in]{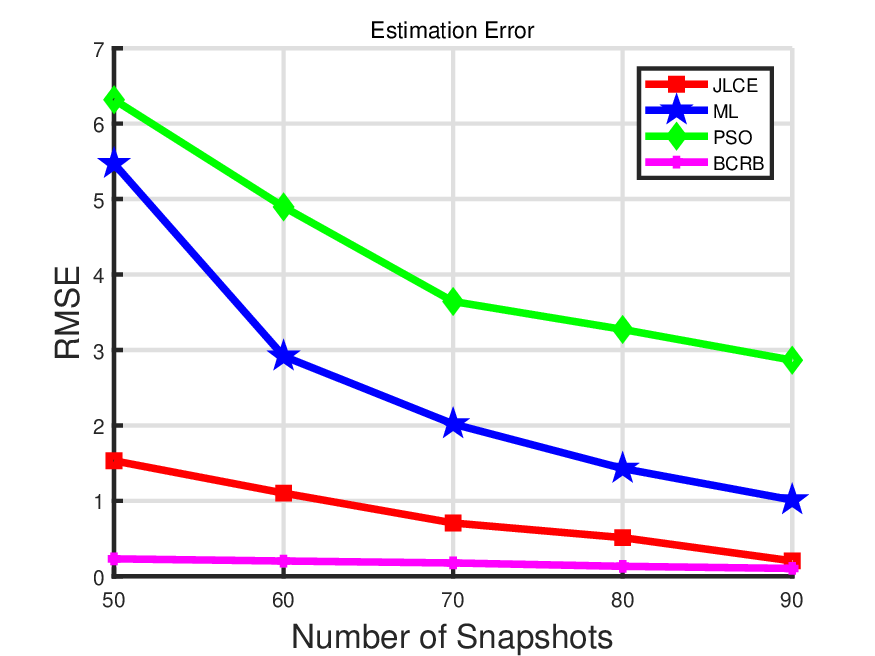}\\
  \caption{Localization performances with  $\text{SNR}=15$ dB and $L=128$ under different snapshots }\label{fig:FigforRIST}
    \vspace{-2em}
\end{figure}

\begin{figure}
  \centering
  \includegraphics[width=3.0in]{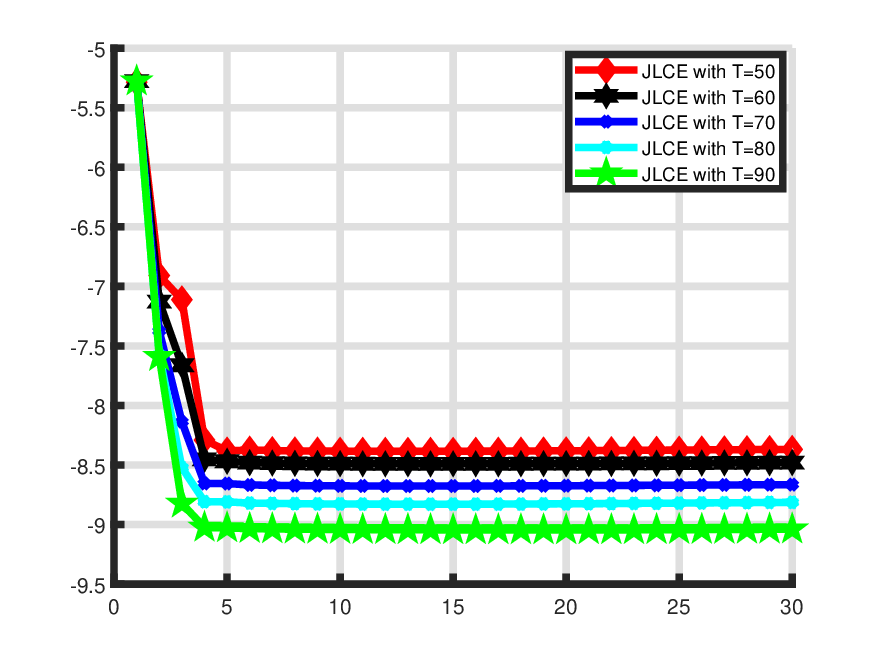}\\
  \caption{Sparse channel vector $\boldsymbol \Delta_{ru}$ estimation results with $\text{SNR}=15 $ dB and $L=128$ under different snapshots }\label{fig:FigforRIST2}
\end{figure}

In Fig.\ref{fig:FigforRIST} and Fig.\ref{fig:FigforRIST2}, the joint localization and channel estimation problem with different snapshot numbers is investigated. The results in Fig.\ref{fig:FigforRIST} and Fig.\ref{fig:FigforRIST2} also support similar conclusions that demonstrate the superiority of the proposed algorithm in localization accuracy, estimation performance as well as convergence rate. Besides, the augmentation of the snapshot number with fixed $PQ$ means the ratio $T/PQ$ changes. The ratio approaching $1$ means the matrix ${{\boldsymbol{\Upsilon A}}}$ is becoming a full sampling matrix.


\begin{figure}
  \centering
  \includegraphics[width=3.0in]{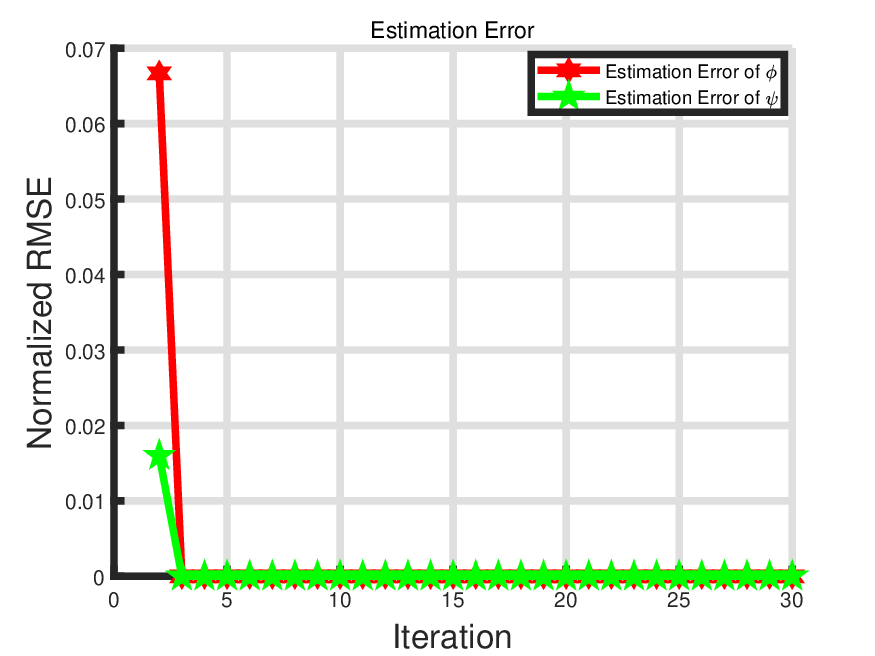}\\
  \caption{Azimuth and elevation angles estimation errors with $\text{SNR}=15 $ dB, $L=128$ and $T=70$ }\label{fig:angle}
\end{figure}

For the channel parameter estimation performances,  the estimation errors can be reduced by increasing the snapshot $T$ or SNR in Fig.\ref{fig:FigforRISSNR2} and Fig.\ref{fig:FigforRIST2}. The vector $\boldsymbol \Delta_{ru}$ is the sparse representation of the reflected path gain $\alpha_{ru}$. The simulation results indicated that the proposed algorithm can precisely estimate the channel parameters simultaneously. Furthermore,  we also investigated the azimuth and elevation angle estimation performances in Fig.\ref{fig:angle}. The simulation results showed the proposed algorithm can accurately estimate the angles in a few iterations.
Meanwhile, the off-grid errors were ignored given enough $PQ$ and it leads to
zero estimation errors in Fig.\ref{fig:angle}. Given the intuitive analysis and simulation results,  the robustness and validity of the proposed algorithm in estimating the channel state information was verified.

\begin{figure}
  \centering
  \includegraphics[width=3.0in]{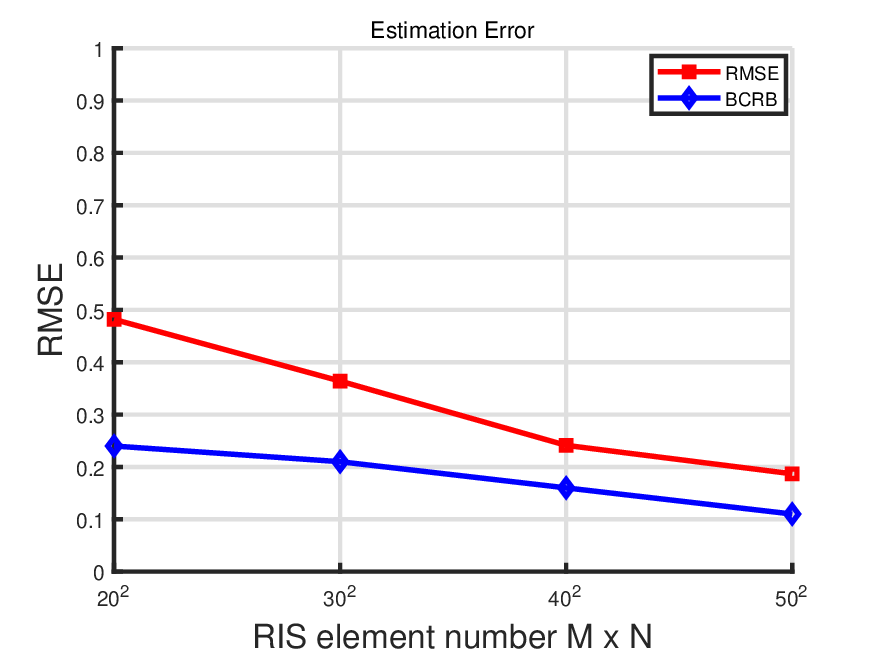}\\
  \caption{{\color{red}Localization performance with $\text{SNR}=15 \text{dB}$, $T=80$ and $L=128$ with different number of RIS elements $M \times N$}}\label{fig:FIgMN1}
\end{figure}

In Fig.\ref{fig:FIgMN1}, the localization performance is investigated under different numbers of RIS elements and other parameters remain unaltered. The number of RIS elements is set to be $M \times N$ with $M=N$. With the augmentation of the RIS element number, the localization accuracy of the JCLE algorithm and the BCRB both decrease and achieve better performance, which demonstrates that the RIS can benefit the localization systems. Moreover, {\color{red}the BCRB and localization error both indicated that there existed a tradeoff between the number of RIS elements and the computational complexity.}

\subsection{\color{red}{Near Field Scenarios}}
 The numerical results of our algorithm in the near-field channel estimation and localization problem are also investigated. The true user location is set to meet the constraint ${\left\| {{{\boldsymbol p}_u} - {{\boldsymbol p}_r}} \right\|_2} \le 100\lambda$ and $\lambda$ is carrier wavelength by following the near-filed settings in \cite{Han22STSPLocalization}.

\begin{figure}
  \centering
  \includegraphics[width=3.0in]{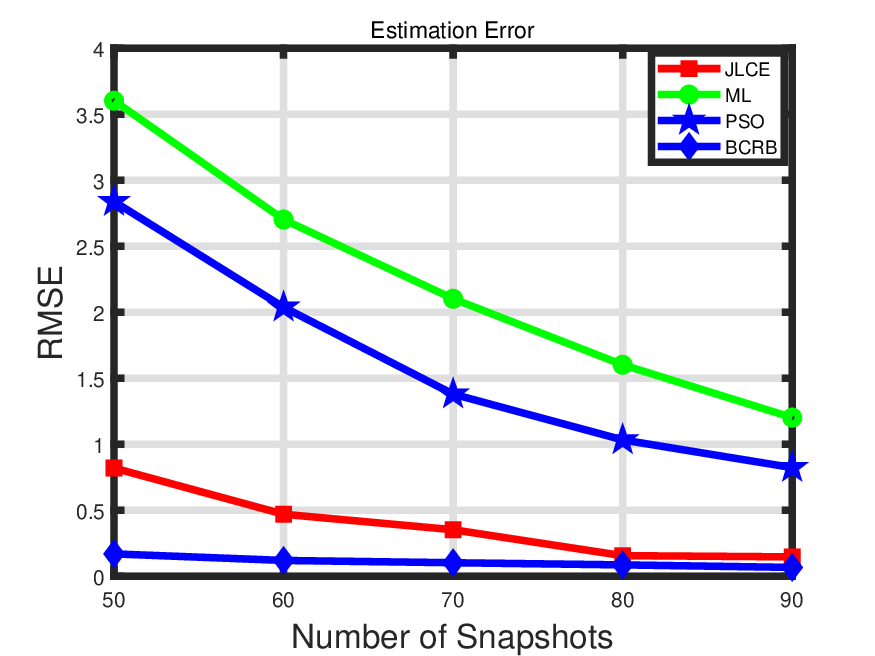}\\
  \caption{Localization performance with  $\text{SNR}=15 \text{dB}$ and $L=128$ under different number of snapshots }\label{fig:nearT1}
\end{figure}

\begin{figure}
  \centering
  \includegraphics[width=3.0in]{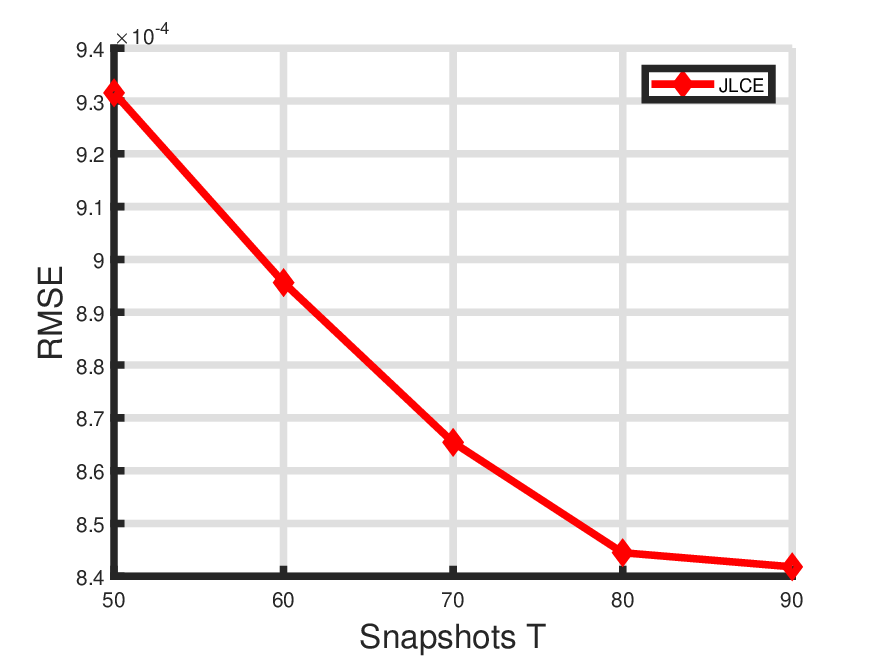}\\
  \caption{Sparse channel vector $\boldsymbol \Delta_{ru}$ estimation results with $\text{SNR}=15 $ dB and  $L=128$ under  different number of snapshots}\label{fig:nearDelta}
\end{figure}

\begin{figure}
  \centering
  \includegraphics[width=3.0in]{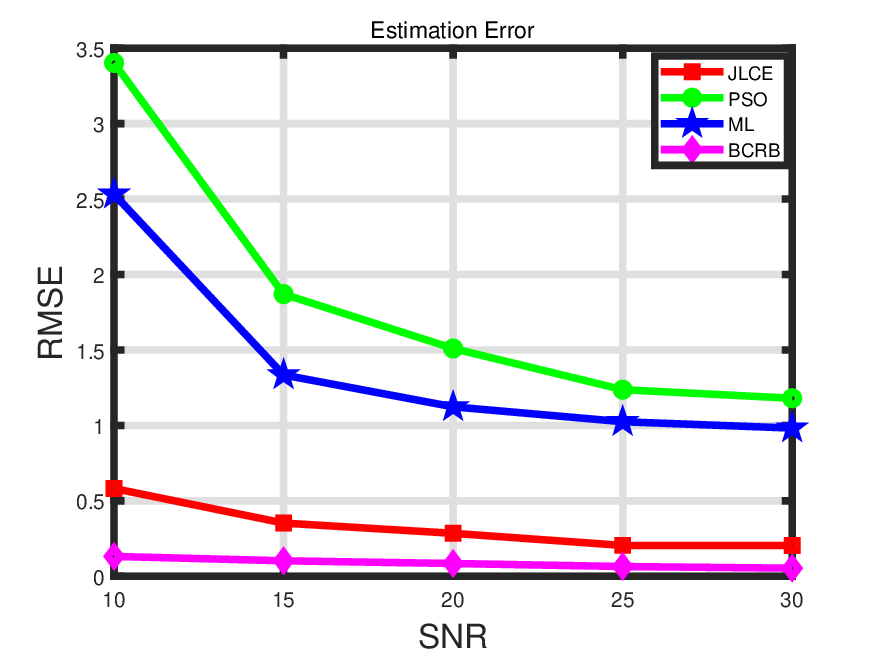}\\
  \caption{Localization performance with  snapshot $T=70$ and $L=128$ with different SNRs }\label{fig:nearT}
\end{figure}

 The simulation results in Fig.\ref{fig:nearT1} and Fig.\ref{fig:nearT}  present the localization performances of the proposed algorithm and other algorithms in the near-filed scenarios. The proposed algorithm can also approach the localization accuracy benchmark BCRB and outperform other compared algorithms.  The numerical results in Fig.\ref{fig:nearDelta} show the estimation error of the sparse vector in the near-field scenario, which also shows that the proposed algorithm can achieve accurate estimation of channel parameters.  The results in near-filed and far-filed scenarios both demonstrate the superiority of the proposed algorithm in localization and validity in channel semation.

\section{Conclusion}

In the paper, we considered a joint localization and channel estimation problem in the RIS-aided system and we proposed a JLCE algorithm to study the complicated estimation problem. Due to the intractable direct maximization of the objective function, the true posterior distribution is approximated by a joint variational distribution iteratively.   In the proposed iterative algorithm, we also investigated the algorithm complexity and convergence. Simulation results have shown the superiority of the proposed algorithm in channel estimation and localization accuracy through various simulation examples.

\appendices

\section{}\label{appendix_alpha}

The transformation matrix $\boldsymbol {\mathcal P }= \frac{{\partial {{ \boldsymbol {\cal W}}}}}{{\partial {\boldsymbol \kappa ^T}}}$ can be calculated as
\begin{equation}\label{Pmatrix}
\boldsymbol{\cal P} = \left[ {\begin{array}{*{20}{c}}
{{\boldsymbol{\cal P}_{11}}}&{{\boldsymbol 0}}\\
\boldsymbol 0&{{\boldsymbol{\cal P}_{22}}}
\end{array}} \right] \in {\mathbb{C}^{7 \times \left( {2L + 2PQ + 2} \right)}},
\end{equation}
where ${\boldsymbol {\cal P}_{11}} = \left[ {\begin{array}{*{20}{c}}
{\frac{{\partial {\boldsymbol{\varphi }}^H\left( {{\zeta _{au}}} \right)}}{{\partial {\boldsymbol{p}}_u}}}&{\frac{{\partial {\boldsymbol{\varphi }}^T\left( {{\zeta _{au}}} \right)}}{{\partial {\zeta _{au}}}}}& {{{\boldsymbol{0}}_{1 \times L}}}\\
{\frac{{\partial {\boldsymbol{\varphi }}^H\left( {{\zeta _{ru}}} \right)}}{{\partial {\boldsymbol{p}}_u}}}& {{{\boldsymbol{0}}_{1 \times L}}}&{\frac{{\partial {\boldsymbol{\varphi }}^H\left( {{\zeta _{ru}}} \right)}}{{\partial {\zeta _{ru}}}}}
\end{array}} \right]^T$ and ${{\boldsymbol{P}}_{22}} = {\left[ {\begin{array}{*{20}{c}}
0&1&{{{\boldsymbol{0}}_{1 \times PQ}}}&{{{\boldsymbol{0}}_{1 \times PQ}}}\\
1&0&{{{\boldsymbol{0}}_{1 \times PQ}}}&{{{\boldsymbol{0}}_{1 \times PQ}}}
\end{array}} \right]^T}$.

The term $\frac{{\partial {{\boldsymbol{\Xi }}_t}}}{{\partial {\boldsymbol{W}}}}$ requires the derivative of $\frac{{\partial {{\boldsymbol{\Xi }}_{au}}}}{{\partial {\boldsymbol{\cal W}}}}$ and $\frac{{\partial {\boldsymbol{\Xi }}_{ru}^t}}{{\partial {\boldsymbol{\cal W}}}}$. $\frac{{\partial {{\boldsymbol{\Xi }}_{au}}}}{{\partial {\boldsymbol{\cal W}}}} = \left[ {{\alpha _{au}}\sqrt {{P_w}} {{\boldsymbol{I}}_L},{\boldsymbol{0}}}\right]  \in {\mathbb{C}}^{L \times \left(2L +2PQ+2\right)}$. Similarly $\frac{{\partial {\boldsymbol{\Xi }}_{ru}^t}}{{\partial {\boldsymbol{\cal W}}}} = \left[ {{\boldsymbol{0}},\boldsymbol{ \mathcal M}^t,\frac{{\partial { \boldsymbol{\Xi }}_{ru}^t}}{{\partial \psi }},\frac{{\partial { \boldsymbol{\Xi }}_{ru}^t}}{{\partial \phi }},{\boldsymbol{0}},{\boldsymbol{0}}} \right] \in {\mathbb{C}}^{L \times \left( {2L + 2 + 2PQ} \right)}$ with
 \begin{equation}\label{Mt}
\boldsymbol{ \mathcal M}^t = {{\alpha _{ru}}\sqrt {{P_w}} {{\boldsymbol{a}}^T}\left( {\theta ,\vartheta } \right){\rm{diag}}\left( {{{\boldsymbol{\Omega }}_t}} \right){\boldsymbol{a}}\left( {\psi ,\phi } \right){{\boldsymbol{I}}_L}}.
\end{equation}

Expanding the product term ${ \frac{{\partial {\boldsymbol{\Xi }}_t^H}}{{\partial {{\boldsymbol{\cal W}}^H}}}\frac{{\partial {{\boldsymbol{\Xi }}_t}}}{{\partial {\boldsymbol{\cal W}}}}}$ and substituting $\frac{{\partial {{\boldsymbol{\Xi }}_{au}}}}{{\partial {\boldsymbol{\cal W}}}}$  and $\frac{{\partial {\boldsymbol{\Xi }}_{ru}^t}}{{\partial {\boldsymbol{\cal W}}}}$  into \eqref{newFIM}, we obtain
\begin{equation}\label{Iris}
\boldsymbol J\left( {\boldsymbol{\cal W}} \right) = {\mathbb{E}_{p\left( {{\boldsymbol{R,}}{\boldsymbol{\cal W}}} \right)}}\left[ {\begin{array}{*{20}{c}}
{\alpha _{au}^2{P_w}{{\boldsymbol{I}}_L}}&{{{\boldsymbol{J}}_{ar}}}&{\boldsymbol{0}}\\
{{\boldsymbol{J}}_{ar}^H}&{\boldsymbol{\cal R}}&{\boldsymbol{0}}\\
{\boldsymbol{0}}&{\boldsymbol{0}}&{\boldsymbol{0}}
\end{array}} \right],
\end{equation}
where ${{\boldsymbol{J}}_{ar}} = {\mathbb{E}_{p\left( {{\boldsymbol{R,}}{\boldsymbol{\cal W}}} \right)}}\left({\alpha _{au}}\sqrt {{P_w}} {{\boldsymbol{I}}_L}{\boldsymbol { Q}^t}\right) = {\boldsymbol{\cal G}}{\boldsymbol {\cal Q}^t}$, ${\boldsymbol {Q}^t} =  \left[ {\begin{array}{*{20}{c}}
{\Re \left( {{{\boldsymbol{\cal M}}^t}} \right)}&{\Re \left( {\frac{{\partial {\boldsymbol{\Xi }}_{ru}^t}}{{\partial \psi }}} \right)}&{\Re \left( {\frac{{\partial {\boldsymbol{\Xi }}_{ru}^t}}{{\partial \phi }}} \right)}
\end{array}} \right]$ and  $\boldsymbol {\cal{R}} = {\mathbb{E}_{p\left( {{\boldsymbol{R,}}{\boldsymbol{\cal W}}} \right)}}\left({\left( {{\boldsymbol {\cal Q}^t}} \right)^H}{\boldsymbol {\cal Q}^t}\right)$.

Substituting \eqref{Pmatrix} and \eqref{Iris} into \eqref{newFIM1}, it yields
\begin{equation}\label{Jkappa}
\boldsymbol J\left( \boldsymbol \kappa  \right) = \left[ {\begin{array}{*{20}{c}}
{{\boldsymbol J_{11}}}&{{\boldsymbol J_{12}}}\\
{\boldsymbol J_{12}^T}&{{\boldsymbol J_{22}}}
\end{array}} \right],
\end{equation}
where the submatrices are given by ${\boldsymbol J_{11}} = {\boldsymbol{\cal U}}^H{\boldsymbol{\cal U}}+ {\delta _{{\alpha _{au}}}}\frac{{\partial {{\boldsymbol{\varphi }}}\left( {{\zeta _{ru}}} \right)}}{{\partial {\boldsymbol{p}}_u}}\frac{{\partial {{\boldsymbol{\varphi }}^H}\left( {{\zeta _{ru}}} \right)}}{{\partial {\boldsymbol{p}}_u}}$ with ${\boldsymbol{\cal U}} = {\left( {\boldsymbol {\cal G} + {{\boldsymbol{\cal M}}^t}} \right)^H}\frac{{\partial {{\boldsymbol{\varphi }}^H}\left( {{\zeta _{ru}}} \right)}}{{\partial {\boldsymbol{p}}_u}}$.
\begin{equation}\label{J12}
{{\boldsymbol{J}}_{12}} = \left[ {\begin{array}{*{20}{c}}
{\frac{{\partial {\boldsymbol{\varphi }}^H\left( {{\zeta _{au}}} \right)}}{{\partial {\boldsymbol{p}}_u}}\left( {{ {\boldsymbol {\cal {\tilde G}}}} + {\boldsymbol{\cal G}}{{\left( {{{\boldsymbol{\cal M}}^t}} \right)}^{{H}}}} \right)\frac{{\partial {{\boldsymbol{\varphi }}}\left( {{\zeta _{au}}} \right)}}{{\partial \zeta _{au}}}}\\
{\frac{{\partial {\boldsymbol{\varphi }}^H\left( {{\zeta _{au}}} \right)}}{{\partial {\boldsymbol{p}}_u}}\left( {{ {\boldsymbol {\cal {\tilde G}}}}{{\left( {{{\boldsymbol{\cal M}}^t}} \right)}^{{H}}} + {{\left( {{{\boldsymbol{\cal M}}^t}} \right)}^{{H}}}{{\boldsymbol{\cal M}}^t}} \right)\frac{{\partial {{\boldsymbol{\varphi }}}\left( {{\zeta _{ru}}} \right)}}{{\partial {\zeta _{ru}}}}}\\
{\frac{{\partial {\boldsymbol{\varphi }}^H\left( {{\zeta _{au}}} \right)}}{{\partial {\boldsymbol{p}}_u}}\left( {{\boldsymbol{\cal G}}{{\left( {{{\boldsymbol{\cal M}}^t}} \right)}^{{H}}} + {{\boldsymbol{\cal M}}^t}} \right){{\left( {\frac{{\partial {\boldsymbol{\Xi }}_{ru}^t}}{{\partial \psi }}} \right)}^H}}\\
{\frac{{\partial {\boldsymbol{\varphi }}^H\left( {{\zeta _{au}}} \right)}}{{\partial {\boldsymbol{p}}_u}}\left( {{\boldsymbol{\cal G}}{{\left( {{{\boldsymbol{\cal M}}^t}} \right)}^{{H}}} + {{\boldsymbol{\cal M}}^t}} \right){{\left( {\frac{{\partial {\boldsymbol{\Xi }}_{ru}^t}}{{\partial \phi }}} \right)}^H}}
\end{array}} \right],
\end{equation}
with ${ {\boldsymbol {\cal {\tilde G}}}} = {\boldsymbol{\cal G}}{{\boldsymbol{\cal G}}^H} + {\delta _{{\alpha _{au}}}}{{\boldsymbol{I}}_L}$ and ${{\boldsymbol{J}}_{22}} = {\boldsymbol{\cal F}}{\boldsymbol{\cal D}}{{\boldsymbol{\cal F}}^H}$,
where
\begin{equation}\label{F}
{\boldsymbol{\cal F}}= {\left[ {\begin{array}{*{20}{c}}
{\frac{{\partial {{\boldsymbol{\varphi }}^H}\left( {{\zeta _{au}}} \right)}}{{\partial {\zeta _{au}}}}}&{\frac{{\partial {{\boldsymbol{\varphi }}^H}\left( {{\zeta _{ru}}} \right)}}{{\partial {\zeta _{ru}}}}}&{\frac{{\partial {\boldsymbol{\Xi }}_{ru}^t}}{{\partial \psi }}}&{\frac{{\partial {\boldsymbol{\Xi }}_{ru}^t}}{{\partial \phi }}}\\
{\boldsymbol{0}}&{\boldsymbol{0}}&{\boldsymbol{0}}&{\boldsymbol{0}}
\end{array}} \right]^H},
\end{equation}
and
\begin{equation}\label{D}
{\boldsymbol{\cal D}} = \left[ {\begin{array}{*{20}{c}}
{{\boldsymbol{\cal {\tilde G}}}}&{{\boldsymbol{\cal G}}{{\boldsymbol{\cal M}}^t}}&{\boldsymbol{\cal G}}&{{{\left( {{{\boldsymbol{\cal G}}^t}} \right)}^{{H}}}}\\
{{{\left( {{{\boldsymbol{\cal M}}^t}} \right)}^{{H}}}{\boldsymbol{\cal G}}}&{{{\left( {{{\boldsymbol{\cal M}}^t}} \right)}^{{H}}}{{\boldsymbol {\cal M}}^t}}&{{{\left( {{{\boldsymbol{\cal M}}^t}} \right)}^{{H}}}}&{{{\left( {{{\boldsymbol{\cal M}}^t}} \right)}^{{H}}}}\\
{\boldsymbol{\cal G}}&{{{\boldsymbol{\cal M}}^t}}&{\boldsymbol{1}}&{{{\left( {{{\boldsymbol{\cal M}}^t}} \right)}^{{H}}}}\\
{{\boldsymbol{\cal G}}^t}&{{{\boldsymbol{\cal M}}^t}}&{{{\left( {{{\boldsymbol{\cal M}}^t}} \right)}^{{H}}}}&{\boldsymbol{1}}
\end{array}} \right].
\end{equation}

\ifCLASSOPTIONcaptionsoff
  \newpage
\fi

\bibliographystyle{IEEEtran}
\bibliography{RISLocalization}

\end{document}